\documentclass[5p,times,twocolumn]{elsarticle}
\usepackage{amsmath,amssymb,bm}
\usepackage{graphicx}
\usepackage{rotating}
\usepackage{booktabs,multirow}
\usepackage{algorithm,algorithmic}
\usepackage{xcolor}
\usepackage{xspace}
\usepackage{pifont}
\usepackage[section]{placeins}
\usepackage[hidelinks]{hyperref}

\newcommand{\cmark}{\ding{51}}
\newcommand{\xmark}{\ding{55}}
\biboptions{sort&compress}
\journal{Knowledge-Based Systems}

\begin{document}

\begin{frontmatter}

\title{RouteDiff: Low-Dose CT Denoising via Fidelity--Refinement Routing and Hypergraph State Regulation}

\author[a,c]{Xinyuan Wang}
\author[b]{Ruili Li}
\author[b]{Tianyang Li}
\author[a,c]{Yan Li}
\author[a,c]{Xuepeng Liu\corref{cor1}}
\ead{2619932@dundee.ac.uk}
\cortext[cor1]{Corresponding author.}
\affiliation[a]{organization={College of Medicine and Biological Information Engineering, Northeastern University},
city={Shenyang 110169},
country={China}}
\affiliation[b]{organization={Tohoku University School of Medicine},
city={Sendai 980-8575},
country={Japan}}
\affiliation[c]{organization={University of Dundee},
city={Dundee DD1 4HN},
country={United Kingdom}}

\begin{abstract}
Diffusion models have become a strong prior for low-dose computed tomography (LDCT) restoration. However, existing samplers apply one structure--detail trade-off to the whole image, set the restoration emphasis of each step by the timestep, and derive each next state recursively from the current one, carrying its errors forward. We propose RouteDiff, a state-aware dual-route diffusion framework that routes detail to where it can be trusted, regulates each reverse step by the state it has reached, and re-enters the trajectory from an anchored estimate. Fidelity--Refinement Routing (FRR) equips one shared U-Net with a conservative fidelity route and a detail-oriented refinement route, fuses them pixel-wise with a confidence map, and exposes their disagreement as route evidence. Hypergraph State Regulation (HSR) organizes the current features and this evidence into a state hypergraph, whose soft hyperedges group scattered regions in the same state. Its readout yields an ordered stage judgment that modulates the bottleneck, and its incidence guides feature interaction in the skip connections. Estimate-Anchored Re-entry (EAR) re-projects a route-aware anchor onto the trajectory at a fixed step and hands the new state back to HSR to be judged again. With the ten network evaluations of its baseline, RouteDiff achieves the best peak signal-to-noise ratio (PSNR) and structural similarity index (SSIM) on four datasets with inserted, real and simulated noise (40.95~dB and 95.13\% on Mayo 2020; 45.18~dB and 97.23\% on Mayo 2016). It also generalizes to unseen doses of real scans and yields the highest Dice (80.92\%) in downstream segmentation.
\end{abstract}


\end{frontmatter}

\section{Introduction}
\label{sec:intro}

Computed tomography (CT) is indispensable in clinical diagnosis, but the associated radiation exposure raises concerns about cumulative risk, particularly for patients who undergo repeated scans \cite{smith2009radiation}. Reducing the tube current lowers the dose at the cost of amplified quantum noise and streak artifacts, which obscure low-contrast structures and fine anatomical detail. Restoring routine-dose image quality from low-dose CT (LDCT) acquisitions has therefore become an important problem in medical image analysis.

\begin{figure}[!b]
\centering
\includegraphics[width=\columnwidth]{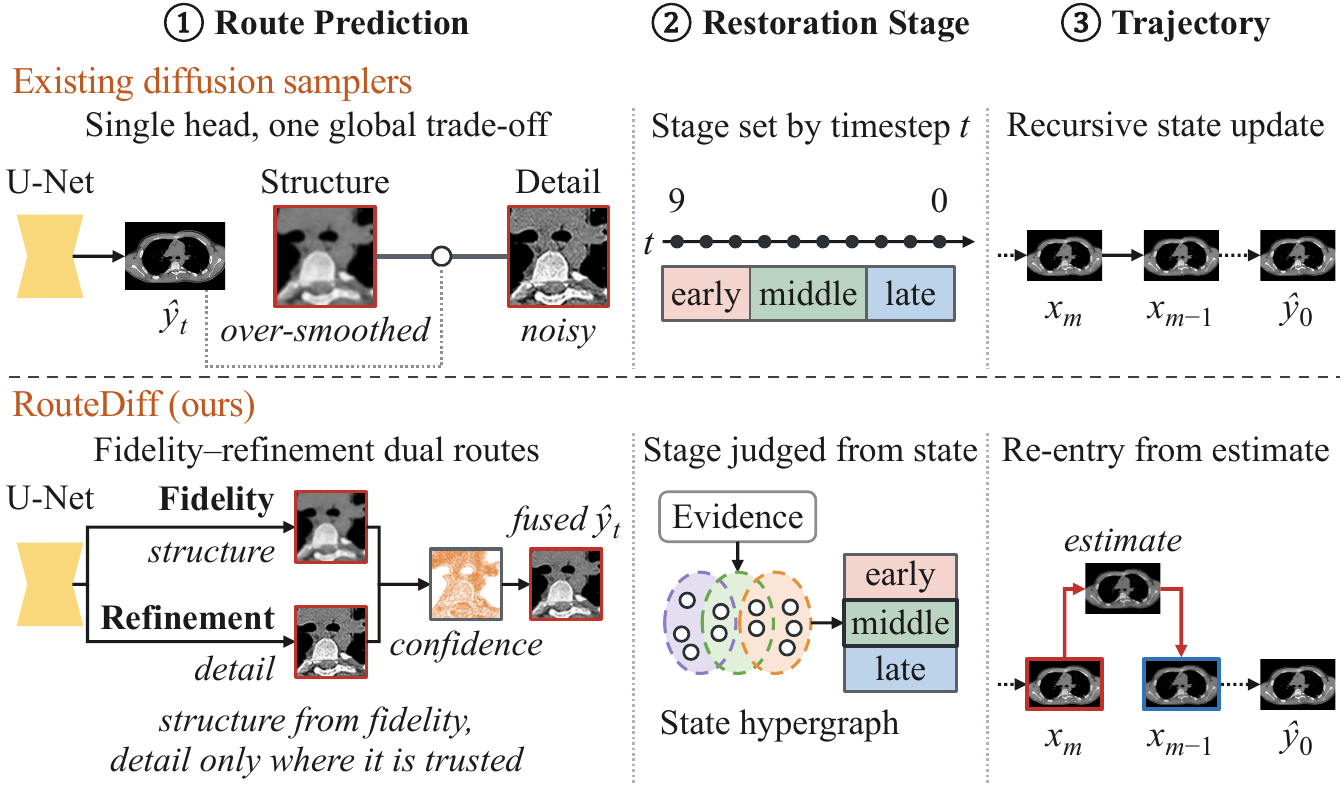}
\caption{Comparison between existing diffusion samplers (top) and RouteDiff (bottom) in route prediction, restoration stage and sampling trajectory.}
\label{fig:intro}
\end{figure}

Deep learning has substantially advanced LDCT restoration. Convolutional encoder--decoder networks \cite{redcnn}, adversarial training \cite{dugan}, and transformer architectures \cite{ctformer} learn a direct mapping from low-dose to routine-dose images. Trained mainly with pixel-wise losses, such mappings tend to regress toward the conditional mean and over-smooth texture and fine structures \cite{kim2019performance}. Diffusion models \cite{ho2020denoising,song2020denoising} provide a powerful generative prior that recovers richer high-frequency detail \cite{i2sb,resshift,rddm}, and generalized or cold diffusion \cite{bansal2023cold} shows that restoration need not start from pure Gaussian noise. For LDCT, interpolating between the low-dose input and the target yields a short trajectory that is consistent with the task \cite{corediff}, and dose- and anatomy-aware conditioning extends diffusion denoising across dose levels and body regions \cite{founddiff}. However, the recovered detail is synthesized, and false structures introduced by a learned prior are a recognized risk in tomographic imaging \cite{bhadra2021hallucinations}. The central difficulty of LDCT restoration is therefore to recover detail without inventing it.

The degradation in LDCT is itself heterogeneous. Automatic tube current modulation varies the dose with the scanned region and the patient size \cite{kalra2004techniques}, so that the noise level and texture differ between slices and between regions. Applied to such data, existing diffusion-based restorers leave three practical problems open (Fig.~\ref{fig:intro}, top).

\begin{figure*}[t]
\centering
\includegraphics[width=\textwidth]{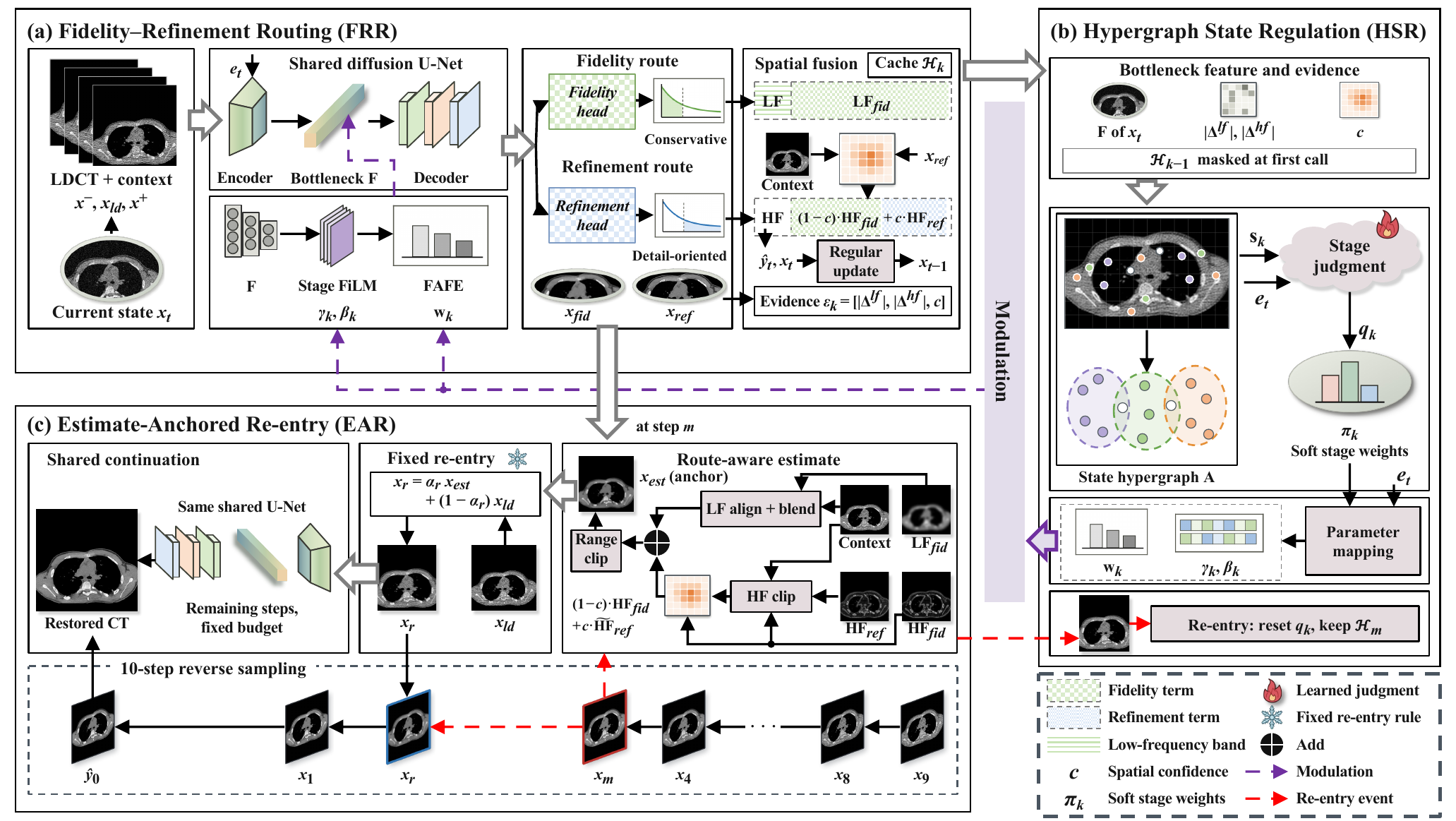}
\caption{Overall framework of RouteDiff. The model consists of a shared diffusion U-Net with (a) Fidelity--Refinement Routing (FRR), (b) Hypergraph State Regulation (HSR) and (c) Estimate-Anchored Re-entry (EAR). The stage FiLM and frequency-aware feature enhancement (FAFE) blocks in (a) apply the modulation computed by HSR, and its skip-connection interaction is shown in Fig.~\ref{fig:srr}(c).}
\label{fig:framework}
\end{figure*}

\emph{Where can generated detail be trusted?} Distortion and perceptual quality are at odds: beyond a certain point, one improves only at the expense of the other \cite{blau2018perception}. Regression networks sit at the low-distortion end, whereas adversarial and diffusion models trade distortion for sharper texture \cite{yi2018sharpness,dugan,corediff}. Predict-and-refine schemes add a stochastic refinement to a deterministic prediction and obtain a better balance \cite{whang2022deblurring}, yet the refinement acts on the whole image and the trade-off remains global. Pixel-wise uncertainty quantification, in contrast, shows that the reliability of a restoration varies strongly across pixels \cite{angelopoulos2022image}.

\emph{How should each step be regulated?} Diffusion restorers decide the restoration emphasis of each step from the timestep, which assumes that all inputs reach the same degradation level at the same step. Even in generic generation, the input of a step at inference comes from the model's own previous predictions rather than from the data, and the resulting exposure bias accumulates along the sampling chain \cite{ning2023input}. Li et al.\ therefore select the timestep from the actual noise level of the sample instead of its nominal index \cite{li2024alleviating}. In LDCT, the mismatch additionally depends on the dose and anatomy of each slice. CoreDiff calibrates the timestep embedding against the input with an error-modulated module \cite{corediff}, and FoundDiff conditions the network on the dose level and anatomical region of the input \cite{founddiff}. Both condition the network on the timestep, corrected or complemented by properties of the input, rather than on an explicit judgment of the restoration stage that the current state has reached. Moreover, the state is not uniform within a slice: homogeneous tissue may be nearly restored while edges and low-contrast structures are not, and the regions that share a state are scattered across the slice rather than forming one neighborhood.

\emph{From which state should sampling continue?} Every reverse step already produces an estimate of the clean image, but in generalized diffusion the next state is a combination of the current state and this estimate \cite{bansal2023cold,corediff}, so errors already present in the state are carried forward \cite{ning2023input}. Several works show that the trajectory need not be followed strictly: CCDF starts the reverse process from a forward-diffused initial estimate and reaches better results in fewer steps \cite{chung2022come}, and Restart sampling re-injects noise midway to contract accumulated errors \cite{xu2023restart}. The former relies on an external estimator and the latter on additional sampling steps, while LDCT diffusion restorers still update their states recursively \cite{corediff,founddiff}.

These problems are coupled: regulating a step by its actual state requires a signal that reflects this state, and continuing from a different state changes the state itself, which then has to be judged again. We therefore propose RouteDiff, a state-aware dual-route diffusion framework that addresses them jointly. It \emph{routes} detail to where it can be trusted, \emph{regulates} each step by the state it has actually reached, and \emph{re-enters} the trajectory from an anchored estimate (Fig.~\ref{fig:intro}, bottom). One shared diffusion U-Net predicts a conservative fidelity route and a detail-oriented refinement route. Low-frequency structure is always taken from the fidelity route, and high-frequency detail is admitted from the refinement route only where a spatial confidence map trusts it. The disagreement between the two routes serves as evidence of the restoration state. Since neither convolution nor pairwise self-attention represents scattered regions of arbitrary size as a whole, we organize the current features and this evidence into a state hypergraph, whose soft hyperedges each connect any number of positions \cite{feng2019hypergraph} and gather the regions in a similar state wherever they lie. Its readout yields an ordered stage judgment that regulates the current step, and its incidence lets regions in the same state exchange information in the skip connections. At a fixed step, the two routes are reused to build a route-aware estimate that is re-projected onto the trajectory, and the re-entered state is handed back to be judged again. The noise schedule and the number of function evaluations (NFE) are unchanged. Fig.~\ref{fig:framework} shows the overall framework.

The main contributions are as follows.
\begin{itemize}
\item We propose Fidelity--Refinement Routing (FRR), which predicts a fidelity and a refinement route with one shared U-Net, routes the high frequencies pixel-wise by a rule-based confidence map that checks the refinement route against the input context, and exposes the route disagreement as evidence.
\item We propose Hypergraph State Regulation (HSR), which judges the restoration stage at every network evaluation from a state hypergraph built on the bottleneck features and the previous route evidence, whose soft hyperedges are defined by learnable state prototypes. The resulting ordered stage judgment modulates the bottleneck in the channel and frequency domains, and the hypergraph incidence drives feature interaction in the skip connections.
\item We propose Estimate-Anchored Re-entry (EAR), which reuses the two routes at a fixed step to construct a route-aware anchor and re-projects it onto the diffusion trajectory without additional NFE. Re-entry resets the stage judgment while keeping the observed evidence.
\item We evaluate RouteDiff on four LDCT datasets with noise inserted into measured projections, real scanner noise and noise simulated from routine-dose images. With ten NFE, it achieves the best peak signal-to-noise ratio (PSNR) and structural similarity index (SSIM) on all four datasets, exceeding the strongest competitors by 0.61~dB on Mayo 2020 and 1.02~dB on Mayo 2016, and yields the highest mean Dice (80.92\%) in downstream organ segmentation. Design analyses verify the main choices, including judging the stage from the state rather than the timestep and grouping regions by a hypergraph rather than by local or pairwise operators.
\end{itemize}

\section{Related work}
\label{sec:related}

\subsection{LDCT restoration and the structure--detail trade-off}
Classical approaches restore LDCT images by sinogram-domain filtering \cite{manduca2009projection} or by iterative reconstruction with statistical priors \cite{xu2012low}. Deep networks have since become the dominant image-domain solution, including residual encoder--decoder networks \cite{redcnn}, adversarial training \cite{dugan}, and transformer-based restorers \cite{ctformer}, while unsupervised frameworks relax the need for paired data \cite{zhang2025sparse}. Pixel-wise losses tend to over-smooth fine structures \cite{kim2019performance}, whereas adversarial objectives sharpen the result but may add spurious texture \cite{yi2018sharpness}. Predict-and-refine schemes add a generative refinement to a deterministic prediction \cite{whang2022deblurring}, and all-in-one medical restorers rely on task-adaptive routing \cite{amir} or frequency-enhanced designs \cite{lu2026allfreqnet}. These methods set the structure--detail trade-off for the whole image or task, whereas FRR routes detail between two output routes pixel-wise.

\subsection{Diffusion-based restoration and sampling}
Denoising diffusion probabilistic models \cite{ho2020denoising} and their deterministic samplers \cite{song2020denoising} provide a strong generative prior for image restoration \cite{xie2025reffusion}. Several works replace the Gaussian endpoint with the degraded observation through bridge formulations \cite{i2sb}, residual shifting \cite{resshift,rddm} or generalized degradations \cite{bansal2023cold}. For LDCT, CoreDiff \cite{corediff} interpolates between the low-dose image and the target and calibrates the timestep embedding with an error-modulated module, and FoundDiff \cite{founddiff} conditions a residual diffusion model on dose- and anatomy-aware embeddings. In generic sampling, exposure bias has been reduced by perturbing the training inputs \cite{ning2023input} or by shifting the sampling timestep to the observed noise level \cite{li2024alleviating}. Other works change the state from which sampling continues: CCDF \cite{chung2022come} starts the reverse process from a forward-diffused estimate, Restart sampling \cite{xu2023restart} alternates forward noise injection with backward integration, and SDEdit \cite{meng2022sdedit} diffuses a guide image before denoising it. In contrast, HSR judges the restoration stage from the bottleneck features and the route evidence, and EAR re-enters the trajectory once from a route-aware anchor built from the model's own routes, without additional NFE.

\subsection{Adaptive conditioning and hypergraph learning}
Feature-wise linear modulation (FiLM) \cite{perez2018film} conditions a network through channel-wise affine transforms. Hypergraph neural networks \cite{feng2019hypergraph} generalize graph convolution to high-order relations, in which a hyperedge may connect any number of nodes, and have been applied to hyperspectral image super-resolution \cite{chen2025hypergraph}. Among vision backbones, Vision GNN \cite{han2022vig} connects image patches by $K$-nearest-neighbor ($K$-NN) graphs, Vision HGNN \cite{han2023vihgnn} groups them into hyperedges by fuzzy $C$-means clustering, and global reasoning units \cite{chen2019glore} project pixel features onto a few latent nodes for global interaction. In HSR, the hyperedges are instead defined by the restoration state, and their readout serves as the state signature.

\section{Method}
\label{sec:method}

\subsection{Overall architecture}
\label{sec:overview}

As shown in Fig.~\ref{fig:framework}, RouteDiff extends a shared diffusion U-Net $\Phi_\theta$ \cite{ronneberger2015unet} with FRR, HSR and EAR, and samples along a generalized diffusion trajectory \cite{bansal2023cold,corediff}. Let $x_{ld}\in[0,1]^{H\times W}$ denote a low-dose slice, $x^-$ and $x^+$ its adjacent slices, and $y$ the routine-dose target. The trajectory interpolates between the target and the low-dose input:
\begin{equation}
x_t=\alpha_t\,y+(1-\alpha_t)\,x_{ld},\qquad \alpha_t=1-t/9
\label{eq:forward}
\end{equation}
where $t=0,1,\dots,9$, so that $x_9=x_{ld}$ and $x_0=y$. Restoration starts from $x_9$ and performs $T=10$ NFE, referred to as calls and indexed by $k=1,\dots,T$, where call $k$ processes step $t=T-k$ at inference. Each call predicts the routine-dose image as $\hat{y}_t$, and for $t>0$ the next state follows the regular update of CoreDiff \cite{corediff}:
\begin{equation}
x_{t-1}=x_t+\frac{\alpha_{t-1}-\alpha_t}{1-\alpha_t}\bigl(\hat{y}_t-x_t\bigr)
\label{eq:update}
\end{equation}
which re-evaluates Eq.~\eqref{eq:forward} at step $t-1$ with $\hat{y}_t$ and the low-dose endpoint $(x_t-\alpha_t\hat{y}_t)/(1-\alpha_t)$ implied by $x_t$, as in the released implementation of CoreDiff. When $x_t=\alpha_t\hat{y}_t+(1-\alpha_t)\,x_{ld}$, this endpoint equals $x_{ld}$, and Eq.~\eqref{eq:update} coincides with the improved sampling of cold diffusion \cite{bansal2023cold}, $x_{t-1}=x_t+(\alpha_{t-1}-\alpha_t)(\hat{y}_t-x_{ld})$.

Within call $k$, the encoder receives $[x_t,x_{ld},x^-,x^+]$ and the timestep embedding $e_t$. It produces the skip features $\mathbf{F}_l$ of each level $l$ and the bottleneck feature $\mathbf{F}\in\mathbb{R}^{C\times H'\times W'}$. Subsequently, HSR judges the restoration stage from $\mathbf{F}$ and the route evidence $\mathcal{H}_{k-1}$ of the previous call, and regulates the bottleneck and the skip connections. The decoder then predicts a fidelity route and a refinement route, which FRR fuses pixel-wise into $\hat{y}_t$ while recording their disagreement as the new evidence $\mathcal{H}_k$. At the fixed step $m$, EAR replaces Eq.~\eqref{eq:update} by re-projecting a route-aware estimate onto the trajectory, and hands the re-entered state back to HSR to be judged again. Finally, the prediction of the last call is the restored image. Since HSR reads the evidence written by FRR, FRR is described first, followed by HSR and EAR.

\subsection{Fidelity--Refinement Routing (FRR)}
\label{sec:drd}

Generated high-frequency texture is trustworthy in some regions and indistinguishable from noise in others, so one structure--detail trade-off cannot serve the whole image. As shown in Fig.~\ref{fig:framework}(a), FRR therefore ends the decoder of $\Phi_\theta$ in two $1\times1$ convolutional heads that share all preceding layers. The fidelity head predicts a conservative estimate $x_{fid}$ and the refinement head a detail-oriented estimate $x_{ref}$. Their roles are induced by how each route is fused and supervised.

The two routes are compared and fused band by band with one fixed frequency decomposition, which HSR and EAR also use. For a map with centered normalized frequency coordinates $(\nu_x,\nu_y)\in[-0.5,0.5]^2$, the normalized frequency $\rho=2\max(|\nu_x|,|\nu_y|)\in[0,1]$ equals one at the Nyquist frequency. The masks $M_L=\mathbf{1}[\rho\le\rho_1]$, $M_M=\mathbf{1}[\rho_1<\rho\le\rho_2]$ and $M_H=\mathbf{1}[\rho>\rho_2]$ select the low, middle and high bands as nested squares of the centered spectrum, where $\mathbf{1}[\cdot]$ is the indicator function, $\rho_1=0.442$ and $\rho_2=0.631$. The low- and high-frequency parts of an image $z$ and the fused prediction are then obtained as follows:
\begin{equation}
\begin{aligned}
z^{lf}&=\mathrm{Re}\,\mathrm{IFFT}\bigl(M_L\odot\mathrm{FFT}(z)\bigr),\qquad z^{hf}=z-z^{lf},\\
\hat{y}_t&=x_{fid}^{lf}+(1-c)\odot x_{fid}^{hf}+c\odot x_{ref}^{hf}
\end{aligned}
\label{eq:fusion}
\end{equation}
where $\odot$ denotes element-wise multiplication, FFT and IFFT denote the two-dimensional fast Fourier transform and its inverse, $\mathrm{Re}$ takes the real part, $z^{hf}$ contains the middle and high bands, and $c\in(0,1]^{H\times W}$ is a confidence map. Low-frequency structure is thus always taken from the fidelity route, whereas high-frequency detail is admitted from the refinement route per pixel by $c$.

Without access to $y$ at inference, $c$ is computed by fixed rules that check whether $x_{ref}$ is consistent with the input context and free of isolated impulsive values. With the context image $x_{ctx}=(x^-+x_{ld}+x^+)/3$, the Sobel gradient magnitude $\nabla_S$ and the $3\times3$ median filter $\mathrm{Med}_3$, three deviations are converted into consistency scores and averaged:
\begin{equation}
\begin{aligned}
d_{edge}&=\bigl|\nabla_S x_{ref}-\nabla_S x_{ctx}\bigr|,\\
d_{ctx}&=\bigl|x_{ref}^{lf}-x_{ctx}^{lf}\bigr|,\\
d_{med}&=\bigl|x_{ref}-\mathrm{Med}_3(x_{ref})\bigr|,\\
c_i&=\exp\bigl(-d_i/d_i^{*}\bigr),\qquad c=\sum\nolimits_{i}w_i\,c_i
\end{aligned}
\label{eq:conf}
\end{equation}
where $i\in\{edge,ctx,med\}$, $w_{edge}=0.4$ and $w_{ctx}=w_{med}=0.3$. The scale $d_i^{*}=\max(Q_{0.9}(d_i),10^{-3})$ is calibrated once before training by computing $d_i$ from $y$ in place of $x_{ref}$ over the training set, where $Q_{0.9}$ is the 90th percentile over all pixels. Each $c_i$ thus approaches zero where its deviation greatly exceeds that of the target. The map is computed under stop-gradient, so that the routes cannot be optimized through their own fusion weights.

Each route is supervised on its contributed bands, and the refinement route also on its discarded low frequencies, so that both estimate the same structure:
\begin{equation}
\begin{aligned}
\ell_{route}=\;&\|x_{fid}^{lf}-y^{lf}\|_1+\|x_{ref}^{lf}-y^{lf}\|_1\\
&+\|(1-c)\odot(x_{fid}^{hf}-y^{hf})\|_1+\|c\odot(x_{ref}^{hf}-y^{hf})\|_1
\end{aligned}
\label{eq:lroute}
\end{equation}

After fusion, FRR records where the routes disagree in structure and in detail, and where detail is trusted. With the route difference $\Delta=x_{ref}-x_{fid}$, the route evidence of call $k$ is defined as:
\begin{equation}
\varepsilon_k=\bigl[\,|\Delta^{lf}|,\ |\Delta^{hf}|,\ c\,\bigr]\in\mathbb{R}^{3\times H\times W}
\label{eq:evidence}
\end{equation}
which requires no ground truth. It is cached as $\mathcal{H}_k=\mathrm{sg}(\varepsilon_k)$ and read by HSR at the next call, where $\mathrm{sg}(\cdot)$ denotes stop-gradient.

\subsection{Hypergraph State Regulation (HSR)}
\label{sec:srr}

\begin{figure*}[t]
\centering
\includegraphics[width=\textwidth]{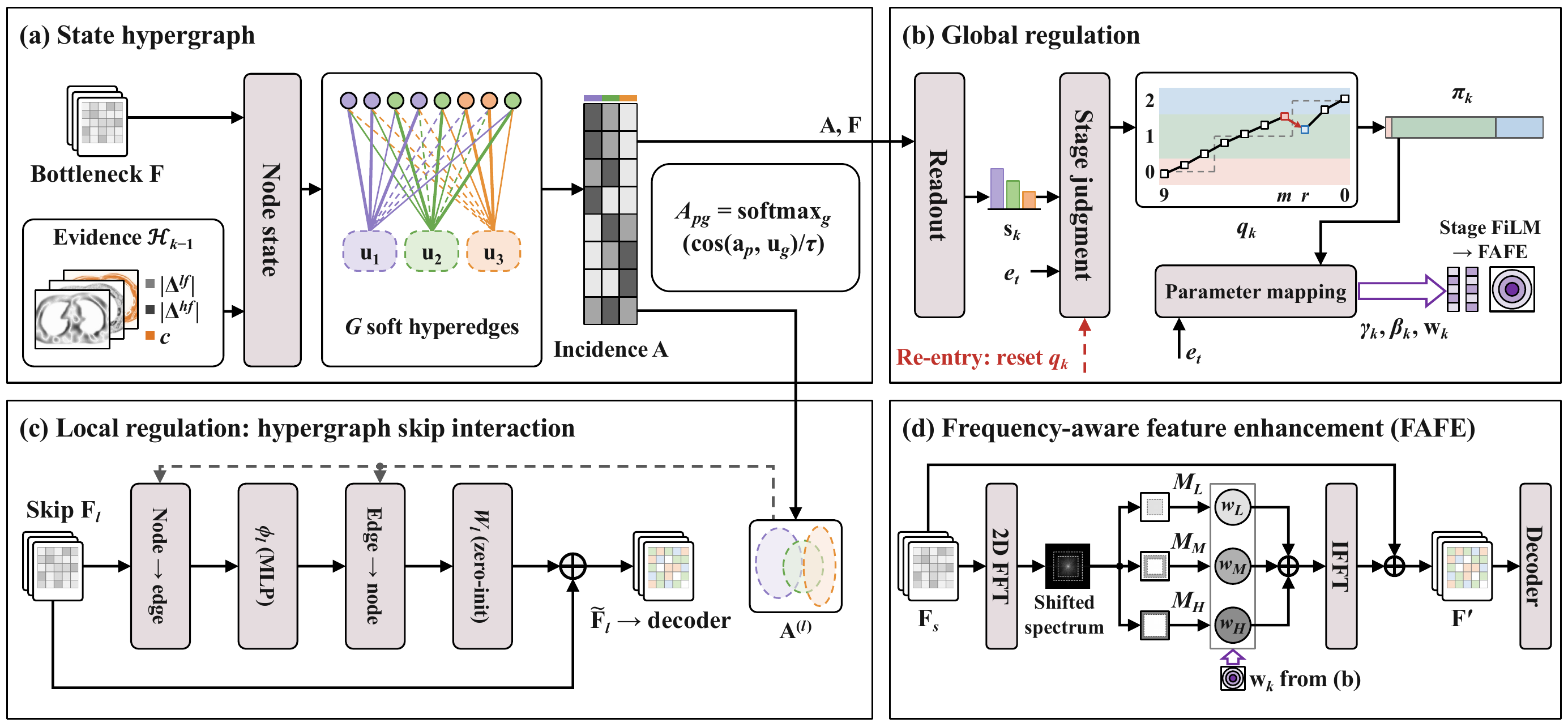}
\caption{Overall framework of Hypergraph State Regulation (HSR). The module consists of (a) state hypergraph construction, (b) global regulation by the stage judgment $q_k$, (c) local regulation by hypergraph interaction in the skip connections and (d) frequency-aware feature enhancement (FAFE).}
\label{fig:srr}
\end{figure*}

The route evidence shows where the routes disagree, but not how far each region has progressed, which the timestep alone cannot reveal. HSR therefore organizes the bottleneck features and the evidence into a state hypergraph (Figs.~\ref{fig:framework}(b) and \ref{fig:srr}), whose hyperedges group regions by state rather than location. Its readout regulates the bottleneck, and its incidence the skip connections.

\subsubsection{State hypergraph construction}
HSR takes the $P=H'W'$ bottleneck positions as nodes (Fig.~\ref{fig:srr}(a)), with features $\mathbf{f}_p\in\mathbb{R}^{C}$ read from $\mathbf{F}$ before modulation. Each node state combines $\mathbf{f}_p$ with the evidence $\bar{\varepsilon}_{k-1}$, average-pooled from $\mathcal{H}_{k-1}$ to the bottleneck resolution, and joins $G$ hyperedges defined by learnable prototypes $\mathbf{u}_g\in\mathbb{R}^{C_a}$:
\begin{equation}
\begin{aligned}
\mathbf{a}_p&=\psi\bigl([\,\mathrm{LN}(\mathbf{f}_p)\,\|\,v_k\,\bar{\varepsilon}_{k-1}(p)\,\|\,v_k\,]\bigr)\in\mathbb{R}^{C_a},\\
A_{pg}&=\mathrm{softmax}_g\bigl(\cos(\mathbf{a}_p,\mathbf{u}_g)/\tau\bigr)
\end{aligned}
\label{eq:node}
\end{equation}
where $\psi$ is a linear layer, $\mathrm{LN}$ is layer normalization, $\|$ denotes concatenation, $\cos(\cdot,\cdot)$ is cosine similarity, $\tau$ is a temperature and $v_k\in\{0,1\}$ indicates valid evidence. At inference, $v_k=0$ only at the first call. The soft incidence $\mathbf{A}=[A_{pg}]\in(0,1)^{P\times G}$ is rebuilt at every call.

\subsubsection{Global regulation}
Subsequently, the normalized degree $n_g$ and the feature $\mathbf{z}_g$ of each hyperedge are read out into the state signature $\mathbf{s}_k$ (Fig.~\ref{fig:srr}(b)):
\begin{equation}
\begin{aligned}
n_g&=\frac{\sum_{p}A_{pg}}{P},\qquad \mathbf{z}_g=\frac{\sum_{p}A_{pg}\,\mathbf{f}_p}{\sum_{p}A_{pg}},\\
\mathbf{s}_k&=\mathrm{MLP}_s\bigl([\,n_1,\dots,n_G\,\|\,W_z\mathbf{z}_1\,\|\cdots\|\,W_z\mathbf{z}_G\,\|\,v_k\,]\bigr)
\end{aligned}
\label{eq:readout}
\end{equation}
where $n_g$ is the share of the slice in hyperedge $g$, $W_z$ is a shared projection and $\mathrm{MLP}_s$ is a multilayer perceptron (MLP). From $\mathbf{s}_k$, HSR derives the stage judgment $q_k\in[0,2]$, whose values $0$, $1$ and $2$ denote the early (structure), middle (balance) and late (detail) stages. Since restoration does not move backward within the segments $9,\dots,m$ and $m-1,\dots,0$ separated by re-entry (Section~\ref{sec:egr}), $q_k$ is initialized at the first call of each segment and only advances:
\begin{equation}
q_k=\begin{cases}2\,\sigma\bigl(f_{init}([\mathbf{s}_k\,\|\,e_t])\bigr), & k\in\mathcal{K}_0\\[2pt] q_{k-1}+(2-q_{k-1})\,\sigma\bigl(f_{upd}([\mathbf{s}_k\,\|\,e_t\,\|\,q_{k-1}])\bigr), & k\notin\mathcal{K}_0\end{cases}
\label{eq:progress}
\end{equation}
where $f_{init}$ and $f_{upd}$ are two-layer MLPs, $\sigma$ is the sigmoid function and $\mathcal{K}_0$ contains these first calls.

On this basis, $q_k$ sets soft stage weights $\pi_k=[\pi_{k,0},\pi_{k,1},\pi_{k,2}]^\top$, and thus the modulation parameters:
\begin{equation}
\begin{aligned}
\pi_{k,j}&=\frac{\exp\bigl(-(q_k-j)^2/\kappa\bigr)}{\sum_{j'=0}^{2}\exp\bigl(-(q_k-j')^2/\kappa\bigr)},\quad j=0,1,2,\\
\gamma_k&=1+\lambda_{mod}\tanh(W_\gamma\pi_k),\qquad \beta_k=\lambda_{mod}\tanh(W_\beta\pi_k),\\
\mathbf{w}_k&=\mathrm{softmax}\bigl(\mathrm{MLP}_w(e_t)+B\pi_k\bigr)
\end{aligned}
\label{eq:mapping}
\end{equation}
where $\kappa$ sets the kernel width, $W_\gamma,W_\beta\in\mathbb{R}^{C\times3}$ are zero-initialized, $\lambda_{mod}$ bounds the modulation and the $j$-th column of $B\in\mathbb{R}^{3\times3}$ is the band bias of stage $j$. Finally, stage FiLM \cite{perez2018film} modulates the bottleneck channel-wise, and frequency-aware feature enhancement (FAFE, Fig.~\ref{fig:srr}(d)) reweights its bands into the decoder input $\mathbf{F}'$:
\begin{equation}
\begin{aligned}
\mathbf{F}_s&=\gamma_k\odot\mathbf{F}+\beta_k,\\
\mathbf{F}'&=\mathbf{F}_s+\mathrm{Re}\,\mathrm{IFFT}\Bigl(\sum\nolimits_{b\in\{L,M,H\}}w_{k,b}\,M_b\odot\mathrm{FFT}(\mathbf{F}_s)\Bigr)
\end{aligned}
\label{eq:filmfafe}
\end{equation}
where $w_{k,b}$ is the entry of $\mathbf{w}_k$ for band $b$, and the FFT is applied per channel with the masks $M_b$ of Section~\ref{sec:drd} on the $H'\times W'$ grid.

\subsubsection{Local regulation}
The regulated bottleneck sets a single emphasis per slice, whereas detail in the skip connections needs region-dependent interaction. HSR therefore reuses $\mathbf{A}$ in all skip connections except the highest-resolution one (Fig.~\ref{fig:srr}(c)). For level $l$, the $G$ membership maps of $\mathbf{A}$ are bilinearly upsampled to its $P_l$ positions, giving $\mathbf{A}^{(l)}$ with memberships summing to one. The features $\mathbf{z}^{(l)}_g$, computed from $\mathbf{A}^{(l)}$ and $\mathbf{F}_l$ as in Eq.~\eqref{eq:readout}, are propagated back:
\begin{equation}
\tilde{\mathbf{F}}_l(p)=\mathbf{F}_l(p)+W_l\Bigl(\sum\nolimits_{g}A^{(l)}_{pg}\,\phi_l\bigl(\mathbf{z}^{(l)}_g\bigr)\Bigr)
\label{eq:skip}
\end{equation}
where $\phi_l$ is a two-layer MLP and $W_l$ is a zero-initialized $1\times1$ convolution, so that training starts from the plain skip connection. Since $G$ is fixed, Eq.~\eqref{eq:skip} costs linear time in the pixel count. With a linear $\phi_l$, it reduces to a hypergraph convolution \cite{feng2019hypergraph} with soft incidence, unit node degrees and unit hyperedge weights. Distant regions in the same state thus share statistics. Boundaries on which the routes disagree carry different evidence and fall mainly into other hyperedges, so they are not smoothed toward homogeneous tissue (Section~\ref{sec:hyper}).

\subsection{Estimate-Anchored Re-entry (EAR)}
\label{sec:egr}

HSR adapts each call to the state it receives, but Eq.~\eqref{eq:update} still rests every state on its predecessor and carries its errors forward. For $m=3$, it gives $x_2=(2x_3+\hat{y}_3)/3$, although after seven of the ten calls the fidelity route already provides reliable structure and the refinement route trusted detail. As shown in Fig.~\ref{fig:framework}(c), EAR therefore builds a route-aware estimate $x_{est}$ from the two routes once at step $m$ and re-projects it onto the trajectory.

Since the re-entered state depends on the network only through $x_{est}$, the estimate applies two context-based corrections to the fused prediction $\hat{y}_m$ of Eq.~\eqref{eq:fusion}. The low frequencies of the fidelity route are aligned with the context. The high frequencies of the refinement route are clipped where neither the context nor the fidelity route supports them:
\begin{equation}
\begin{aligned}
\tilde{x}_{fid}^{lf}&=\lambda_{lf}\,\mathrm{Align}\bigl(x_{fid}^{lf},x_{ctx}^{lf}\bigr)+(1-\lambda_{lf})\,x_{ctx}^{lf},\\
\tilde{x}_{ref}^{hf}&=\mathrm{clip}\bigl(x_{ref}^{hf},-\delta\eta,\delta\eta\bigr)
\end{aligned}
\label{eq:estparts}
\end{equation}
where $\lambda_{lf}=0.7$ weights the aligned structure and $\mathrm{Align}(z,z')$ matches the per-image mean and standard deviation of $z$ to those of $z'$. The bound $\delta\eta$ combines $\delta=2$ with the pixel-wise high-frequency support $\eta=\max\bigl(|x_{ctx}^{hf}|,|x_{fid}^{hf}|\bigr)$. On this basis, the corrected bands are fused with $c$ into the anchor of re-entry:
\begin{equation}
x_{est}=\mathrm{clip}\bigl(\tilde{x}_{fid}^{lf}+(1-c)\odot x_{fid}^{hf}+c\odot\tilde{x}_{ref}^{hf},\,0,\,1\bigr)
\label{eq:xest}
\end{equation}

Subsequently, the anchor is re-projected to step $r=m-1$ by Eq.~\eqref{eq:forward} with $x_{est}$ in place of $y$: \begin{equation} x_r=\alpha_r\,x_{est}+(1-\alpha_r)\,x_{ld} \label{eq:reentry} \end{equation} where $\alpha_2=7/9$ for $m=3$. This state replaces the regular update at step $m$. The re-entry step and $\alpha_r$ are fixed rather than decided by HSR, and no NFE is added, since time still advances from $m$ to $r$.

Re-entry builds $x_r$ from the new anchor $x_{est}$ instead of its predecessor, so the stage reached by the previous calls no longer applies to it. The stage judgment is therefore reset, and the call at step $r$ starts a new segment of Eq.~\eqref{eq:progress}. The evidence $\mathcal{H}_{T-m}$ written at step $m$, denoted $\mathcal{H}_m$ for brevity, still records where the routes disagreed. It is therefore kept with $v_k=1$, so that the state hypergraph of $x_r$ and the new judgment can refer to it. Algorithm~\ref{alg:infer} summarizes the inference procedure.

\begin{algorithm}[t]
\caption{Inference procedure of RouteDiff}
\label{alg:infer}
\begin{algorithmic}[1]
\REQUIRE Low-dose slice $x_{ld}$, neighbors $x^-,x^+$, re-entry step $m$
\ENSURE Restored image $\hat{y}_0$
\STATE $x_9\gets x_{ld}$, $\mathcal{H}_0\gets\mathbf{0}$, $\mathcal{K}_0\gets\{1,T-m+1\}$
\FOR{$k=1$ \TO $T$}
\STATE $t\gets T-k$
\STATE Encode $[x_t,x_{ld},x^-,x^+]$ and $e_t$ into $\{\mathbf{F}_l\}$ and $\mathbf{F}$
\STATE Build $\mathbf{A}$ and $\mathbf{s}_k$ from $\mathbf{F}$ and $\mathcal{H}_{k-1}$ \COMMENT{Eqs.~\eqref{eq:node}--\eqref{eq:readout}}
\STATE Compute $q_k$ and $\pi_k$ \COMMENT{Eqs.~\eqref{eq:progress}, \eqref{eq:mapping}}
\STATE Modulate $\mathbf{F}$, $\{\mathbf{F}_l\}$; decode $x_{fid}$, $x_{ref}$ \COMMENT{Eqs.~\eqref{eq:mapping}--\eqref{eq:skip}}
\STATE Fuse $x_{fid}$ and $x_{ref}$ into $\hat{y}_t$ \COMMENT{Eqs.~\eqref{eq:fusion}--\eqref{eq:conf}}
\STATE $\mathcal{H}_k\gets\mathrm{sg}(\varepsilon_k)$ \COMMENT{Eq.~\eqref{eq:evidence}}
\IF{$t=m$}
\STATE Build the anchor $x_{est}$ \COMMENT{Eqs.~\eqref{eq:estparts}, \eqref{eq:xest}}
\STATE $x_{t-1}\gets\alpha_{t-1}\,x_{est}+(1-\alpha_{t-1})\,x_{ld}$ \COMMENT{Eq.~\eqref{eq:reentry}}
\ELSIF{$t>0$}
\STATE $x_{t-1}\gets x_t+\frac{\alpha_{t-1}-\alpha_t}{1-\alpha_t}(\hat{y}_t-x_t)$ \COMMENT{Eq.~\eqref{eq:update}}
\ENDIF
\ENDFOR
\end{algorithmic}
\end{algorithm}

\subsection{Training}
\label{sec:training}

Since HSR reads the evidence of the previous call and judges re-entered states, training unrolls short call sequences of the same network. Each iteration samples one of two branches, whose first call processes a state of Eq.~\eqref{eq:forward} as the start of a segment with $v_k=0$. The regular branch executes calls at steps $t+1$ and $t$, and its second call processes the regularly updated $x_t$ with the evidence of the first. Here, $t\neq m-1$, since the transition from $m$ to $m-1$ is always a re-entry at inference. The re-entry branch executes calls at steps $m+1$, $m$ and $r$. Its second call processes the regularly updated $x_m$ and provides $x_{est}$, and its third call processes $x_r$ as a new segment with the evidence of the second. Image states and evidence passed between calls are detached, whereas $q_k$ keeps its gradient so that the ordered update of Eq.~\eqref{eq:progress} can be learned. The objective over the set $\mathcal{J}$ of calls in an iteration is:
\begin{equation}
\begin{aligned}
\mathcal{L}&=\sum\nolimits_{k\in\mathcal{J}}\omega_k\,\mathcal{L}_k,\\
\mathcal{L}_k&=\|\hat{y}_{t_k}-y\|_2^2+\ell_{route,k}+\lambda_{pri}\bigl(q_k-\bar{q}(t_k)\bigr)^2
\end{aligned}
\label{eq:loss}
\end{equation}
where $t_k$ is the step of call $k$, $\ell_{route,k}$ denotes Eq.~\eqref{eq:lroute} at call $k$, $\omega_k=1$ for the re-entered call and $1/2$ otherwise, and $\lambda_{pri}=0.01$ weights the stage prior $\bar{q}(t)$. This prior assigns $0$, $1$ and $2$ to the early, middle and late steps $\{7,8,9\}$, $\{3,\dots,6\}$ and $\{0,1,2\}$. Used only in training, it ties $q_k$ to the meaning of the three stages. Apart from it, HSR learns only through the reconstruction it helps to produce.

\section{Experiments}
\label{sec:experiments}

We evaluate RouteDiff on four CT datasets that differ in anatomy, dose and noise source: Mayo 2020, which combines noise insertion into measured projections with projection-domain simulation from routine-dose images; Mayo 2016, with noise inserted into measured projections; Piglet, with real scanner noise at seen and unseen doses, used as an external test set for the Mayo 2020 models; and AMOS, with simulated noise and iterative reconstruction, on which downstream organ segmentation is also evaluated.

\begin{figure*}[t]
\centering
\includegraphics[width=\textwidth]{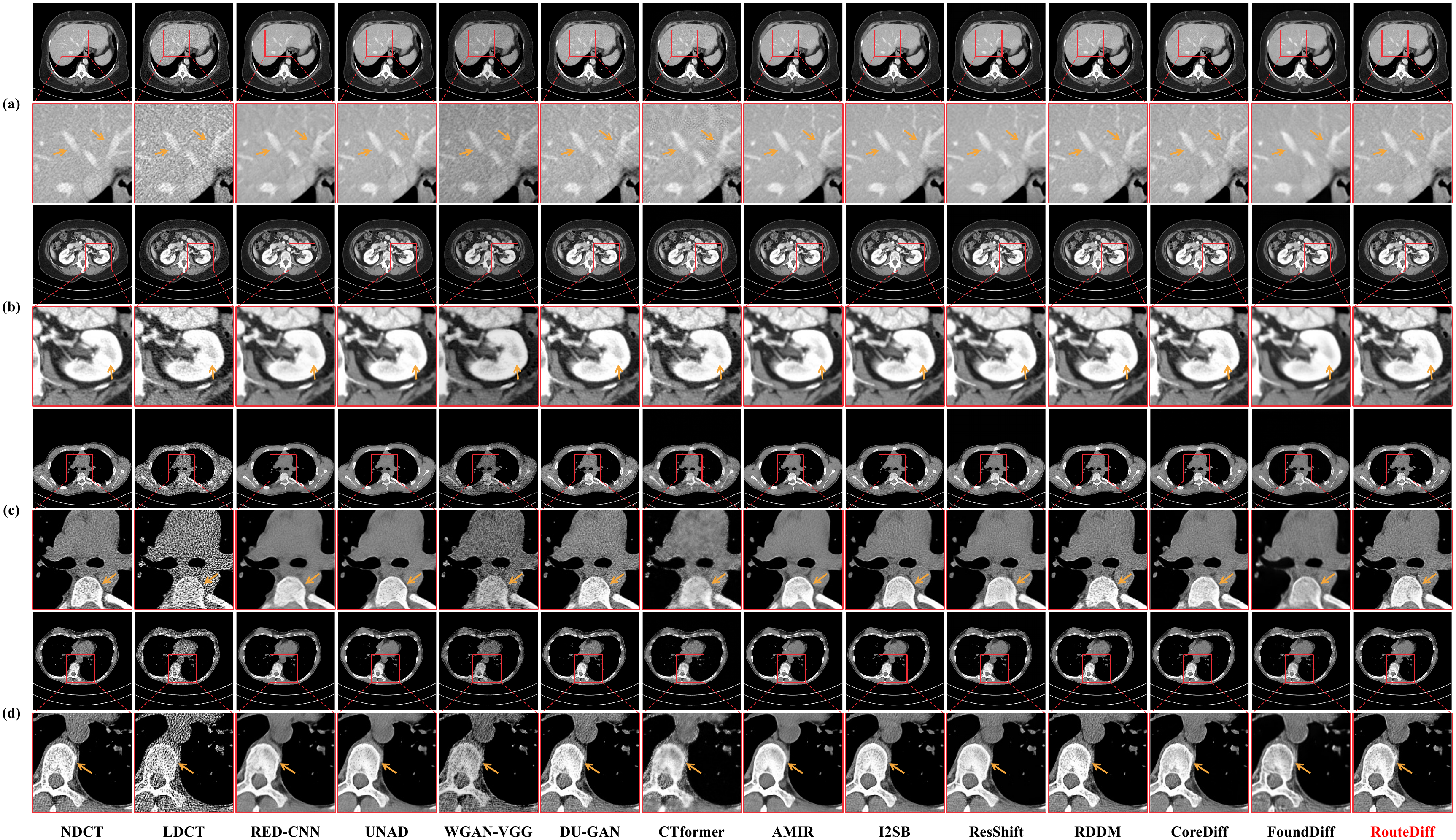}
\caption{Visual results of different methods on the Mayo 2020 dataset. (a), (b) Abdominal slices at $1/4$ dose; (c), (d) chest slices at $1/10$ dose. NDCT denotes the routine-dose reference.}
\label{fig:mayo20}
\end{figure*}

\begin{table*}[t]
\centering
\caption{Comparison results of different methods on the Mayo 2020 dataset (mean $\pm$ standard deviation). The best and second-best results are highlighted in bold and underlined, respectively.}
\label{tab:mayo20}
\footnotesize
\setlength{\tabcolsep}{3pt}
\begin{tabular*}{\textwidth}{@{\extracolsep{\fill}}lccccc}
\toprule
Methods & PSNR (dB)$\uparrow$ & SSIM (\%)$\uparrow$ & FSIM (\%)$\uparrow$ & NQM (dB)$\uparrow$ & VIF$\uparrow$\\
\midrule
RED-CNN & 39.52$\pm$0.36 & 94.34$\pm$0.15 & 93.80$\pm$1.48 & 27.37$\pm$0.33 & 0.550$\pm$0.014\\
UNAD & 39.85$\pm$0.23 & 94.72$\pm$0.08 & 94.35$\pm$1.82 & \underline{28.08$\pm$0.19} & 0.576$\pm$0.018\\
WGAN-VGG & 38.91$\pm$0.40 & 93.28$\pm$0.18 & 94.27$\pm$1.16 & 26.86$\pm$0.44 & 0.536$\pm$0.006\\
DU-GAN & 38.33$\pm$0.31 & 92.60$\pm$0.15 & 94.61$\pm$1.69 & 26.38$\pm$0.25 & 0.522$\pm$0.020\\
CTformer & 35.16$\pm$0.37 & 91.10$\pm$0.20 & 91.61$\pm$1.08 & 21.34$\pm$0.40 & 0.430$\pm$0.011\\
AMIR & 39.82$\pm$0.27 & \underline{94.82$\pm$0.14} & 93.92$\pm$1.91 & \textbf{28.42$\pm$0.18} & \underline{0.581$\pm$0.021}\\
I2SB & 39.72$\pm$0.38 & 94.41$\pm$0.22 & 94.42$\pm$1.33 & 27.41$\pm$0.45 & 0.578$\pm$0.007\\
ResShift & 37.98$\pm$0.32 & 93.34$\pm$0.19 & \textbf{94.97$\pm$1.71} & 26.41$\pm$0.30 & 0.539$\pm$0.022\\
RDDM & 39.82$\pm$0.41 & 94.53$\pm$0.16 & 94.68$\pm$1.19 & 26.92$\pm$0.26 & 0.568$\pm$0.019\\
CoreDiff & 40.26$\pm$0.24 & 94.40$\pm$0.13 & 94.40$\pm$1.61 & 27.26$\pm$0.39 & \underline{0.581$\pm$0.005}\\
FoundDiff & \underline{40.34$\pm$0.39} & 94.29$\pm$0.17 & 94.52$\pm$1.38 & 27.39$\pm$0.35 & 0.577$\pm$0.023\\
\midrule
RouteDiff (ours) & \textbf{40.95$\pm$0.22} & \textbf{95.13$\pm$0.11} & \underline{94.91$\pm$1.32} & 27.81$\pm$0.34 & \textbf{0.584$\pm$0.014}\\
\bottomrule
\end{tabular*}
\end{table*}

\subsection{Datasets}
\label{sec:datasets}

Following CoreDiff \cite{corediff}, all CT images are converted to Hounsfield units (HU), clipped to $[-1024,3072]$~HU and linearly normalized to $[0,1]$, and all metrics are computed within $[-1000,1000]$~HU. All models take three adjacent slices as input context, with the center slice as the restoration target.

Mayo 2020 \cite{mayo2020} provides paired low-dose and routine-dose CT scans with their projection data; its low-dose scans were produced by validated projection-domain noise insertion at $1/4$ dose for the abdomen and $1/10$ dose for the chest. We use 5,151 paired slices from 20 patients covering both anatomies. To cover both doses for each anatomy, we keep the original scans and synthesize the two missing dose--anatomy combinations from the routine-dose images following FoundDiff \cite{founddiff}. Each image is forward-projected with the ASTRA toolbox \cite{vanaarle2016astra}, Poisson noise at a dose-scaled photon count and Gaussian electronic noise are added to the projections, and the result is reconstructed by filtered back projection. This yields 5,151 pairs per dose and 10,302 pairs in total, which share the routine-dose references. The data are split by patient into 16 training and 4 test patients, so that all slices and dose versions of a patient belong to the same subset.

Mayo 2016 \cite{mccollough2017ldgc} is the abdominal CT dataset of the 2016 AAPM Low Dose CT Grand Challenge, whose $1/4$-dose scans were simulated by inserting noise into the routine-dose projection data. It contains 5,936 paired slices from 10 patients, of which 8 patients are used for training and 2 for testing.

Piglet \cite{yi2018sharpness} is a real multi-dose CT dataset of one animal, scanned at 300~mAs (routine dose) and at 150, 75, 30 and 15~mAs, i.e., $1/2$, $1/4$, $1/10$ and $1/20$ of the routine dose, with 850 image pairs per low-dose level. As the low-dose images are separate acquisitions, their noise is produced by the scanner itself. Piglet serves as an external test set: every method is applied with its Mayo 2020 weights, and the 300~mAs scans are used as references. Since Mayo 2020 contains $1/4$ and $1/10$ dose, these two doses are seen during training, whereas $1/2$ and $1/20$ are unseen.

AMOS \cite{ji2022amos} is a large-scale abdominal multi-organ benchmark that provides routine-dose CT scans without paired low-dose acquisitions. We use its 300 annotated CT scans, split them at the patient level into 240 training and 60 test patients, and select five slices per patient, yielding 1,200 training and 300 test slices, all resampled to $512\times512$. Low-dose inputs, including the two context neighbors of each slice, are synthesized with TIGRE \cite{biguri2016tigre}. Each reference image is forward-projected at 360 views with 1,024 detector bins, and Poisson noise with $I_0=5\times10^6$ incident photons and Gaussian noise with a standard deviation of 20 are added in the projection domain. The degraded image is then reconstructed with 100 iterations of ordered-subset simultaneous algebraic reconstruction technique (OS-SART) \cite{wang2004ossart}. The original images serve as targets.

\begin{figure*}[t]
\centering
\includegraphics[width=\textwidth]{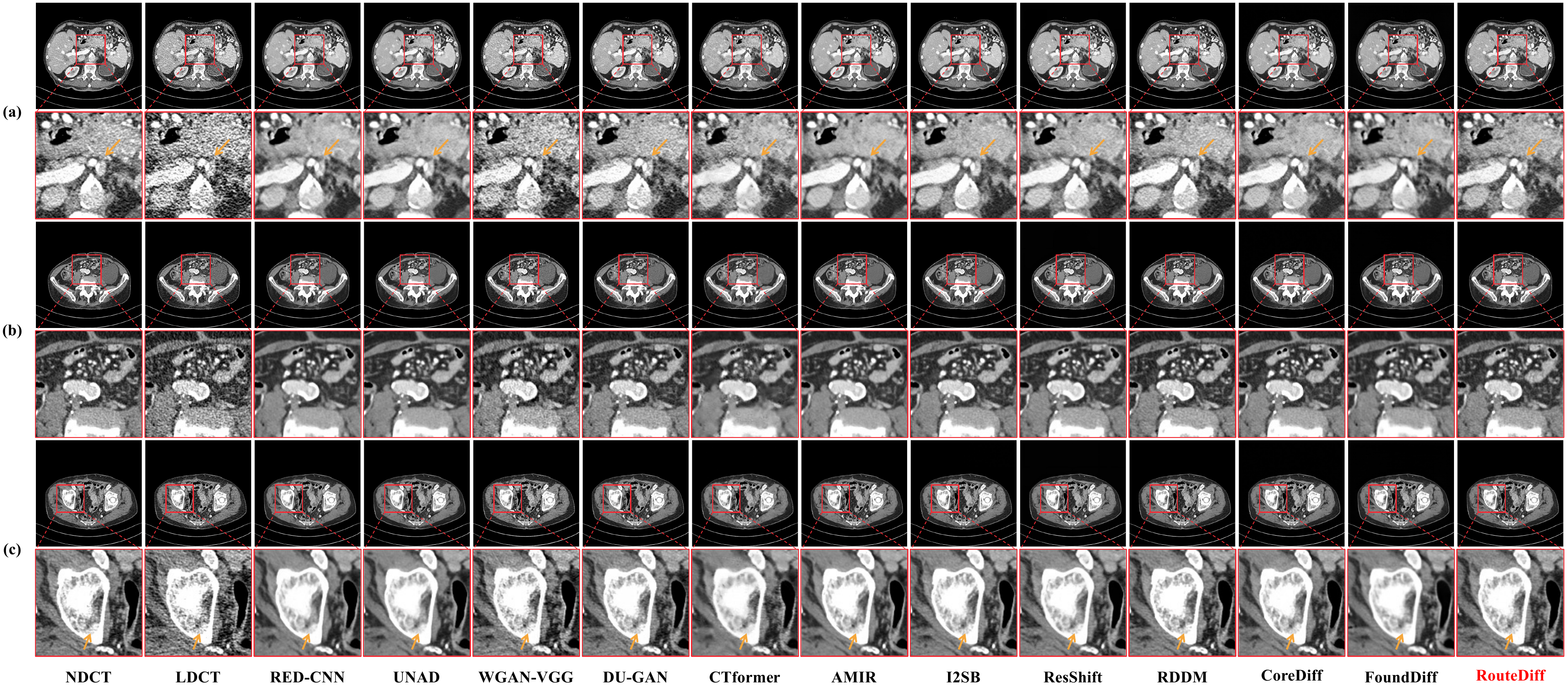}
\caption{Visual results of different methods on the Mayo 2016 dataset.}
\label{fig:mayo16}
\end{figure*}

\begin{table*}[t]
\centering
\caption{Comparison results of different methods on the Mayo 2016 dataset (mean $\pm$ standard deviation). The best and second-best results are highlighted in bold and underlined, respectively.}
\label{tab:mayo16}
\footnotesize
\setlength{\tabcolsep}{3pt}
\begin{tabular*}{\textwidth}{@{\extracolsep{\fill}}lccccc}
\toprule
Methods & PSNR (dB)$\uparrow$ & SSIM (\%)$\uparrow$ & FSIM (\%)$\uparrow$ & NQM (dB)$\uparrow$ & VIF$\uparrow$\\
\midrule
RED-CNN & 43.91$\pm$0.25 & 96.05$\pm$0.07 & 96.48$\pm$1.58 & 31.33$\pm$0.27 & 0.528$\pm$0.013\\
UNAD & 43.90$\pm$0.29 & 95.98$\pm$0.12 & 95.71$\pm$1.65 & 30.89$\pm$0.16 & 0.506$\pm$0.012\\
WGAN-VGG & 40.15$\pm$0.28 & 94.79$\pm$0.06 & 95.36$\pm$0.31 & 29.89$\pm$0.32 & 0.512$\pm$0.009\\
DU-GAN & 40.79$\pm$0.30 & 95.60$\pm$0.11 & 95.14$\pm$1.26 & 28.96$\pm$0.36 & 0.444$\pm$0.015\\
CTformer & 43.11$\pm$0.25 & 92.88$\pm$0.06 & 95.27$\pm$1.21 & 30.26$\pm$0.37 & 0.457$\pm$0.010\\
AMIR & 43.84$\pm$0.33 & 96.46$\pm$0.09 & \underline{97.02$\pm$1.74} & 31.55$\pm$0.24 & 0.519$\pm$0.012\\
I2SB & 43.16$\pm$0.34 & 94.68$\pm$0.13 & 96.63$\pm$1.42 & 30.63$\pm$0.41 & 0.510$\pm$0.016\\
ResShift & 43.41$\pm$0.35 & 95.50$\pm$0.10 & 96.51$\pm$1.44 & 30.25$\pm$0.14 & 0.489$\pm$0.004\\
RDDM & 43.90$\pm$0.25 & 96.34$\pm$0.12 & 96.03$\pm$1.37 & 29.74$\pm$0.48 & 0.519$\pm$0.013\\
CoreDiff & \underline{44.16$\pm$0.26} & 96.36$\pm$0.10 & 97.01$\pm$1.54 & 31.76$\pm$0.22 & 0.526$\pm$0.015\\
FoundDiff & 44.11$\pm$0.35 & \underline{96.74$\pm$0.06} & 96.69$\pm$1.47 & \underline{31.82$\pm$0.31} & \underline{0.530$\pm$0.008}\\
\midrule
RouteDiff (ours) & \textbf{45.18$\pm$0.25} & \textbf{97.23$\pm$0.06} & \textbf{97.37$\pm$1.28} & \textbf{32.41$\pm$0.23} & \textbf{0.533$\pm$0.015}\\
\bottomrule
\end{tabular*}
\end{table*}

\subsection{Implementation details}
\label{sec:implementation}

All experiments are conducted on a single NVIDIA GeForce RTX 5090 GPU. The backbone is the U-Net of CoreDiff \cite{corediff}, extended by the second prediction head and HSR, and a separate model is trained on each of Mayo 2020, Mayo 2016 and AMOS. RouteDiff is trained for 100,000 iterations with the Adam optimizer, an initial learning rate of $2\times10^{-4}$ and a batch size of 4, and the exponential moving average (EMA) of the weights at the final iteration is used for testing. The step $t$ of the second call in the regular branch follows a curriculum that starts from the early steps $\{7,8,9\}$ of the stage prior $\bar{q}$ and gradually includes the middle and late steps. Inference uses ten NFE with re-entry at $m=3$, and the same hyperparameters are used for all datasets.

We compare RouteDiff with RED-CNN \cite{redcnn}, UNAD \cite{unad}, WGAN-VGG \cite{wganvgg}, DU-GAN \cite{dugan}, CTformer \cite{ctformer}, AMIR \cite{amir}, I2SB \cite{i2sb}, ResShift \cite{resshift}, RDDM \cite{rddm}, CoreDiff \cite{corediff} and FoundDiff \cite{founddiff}. Each baseline is retrained from its official implementation on the same splits, preprocessing and three-slice input with its published hyperparameters.

PSNR and SSIM are reported on all datasets. On Mayo 2020 and Mayo 2016, we additionally report the feature similarity index (FSIM) \cite{fsim}, the noise quality measure (NQM) \cite{nqm} and visual information fidelity (VIF) \cite{vif}. All metrics are computed per test slice and reported as their mean, together with the standard deviation ($\pm$) where it is given.

\begin{figure*}[t]
\centering
\includegraphics[width=\textwidth]{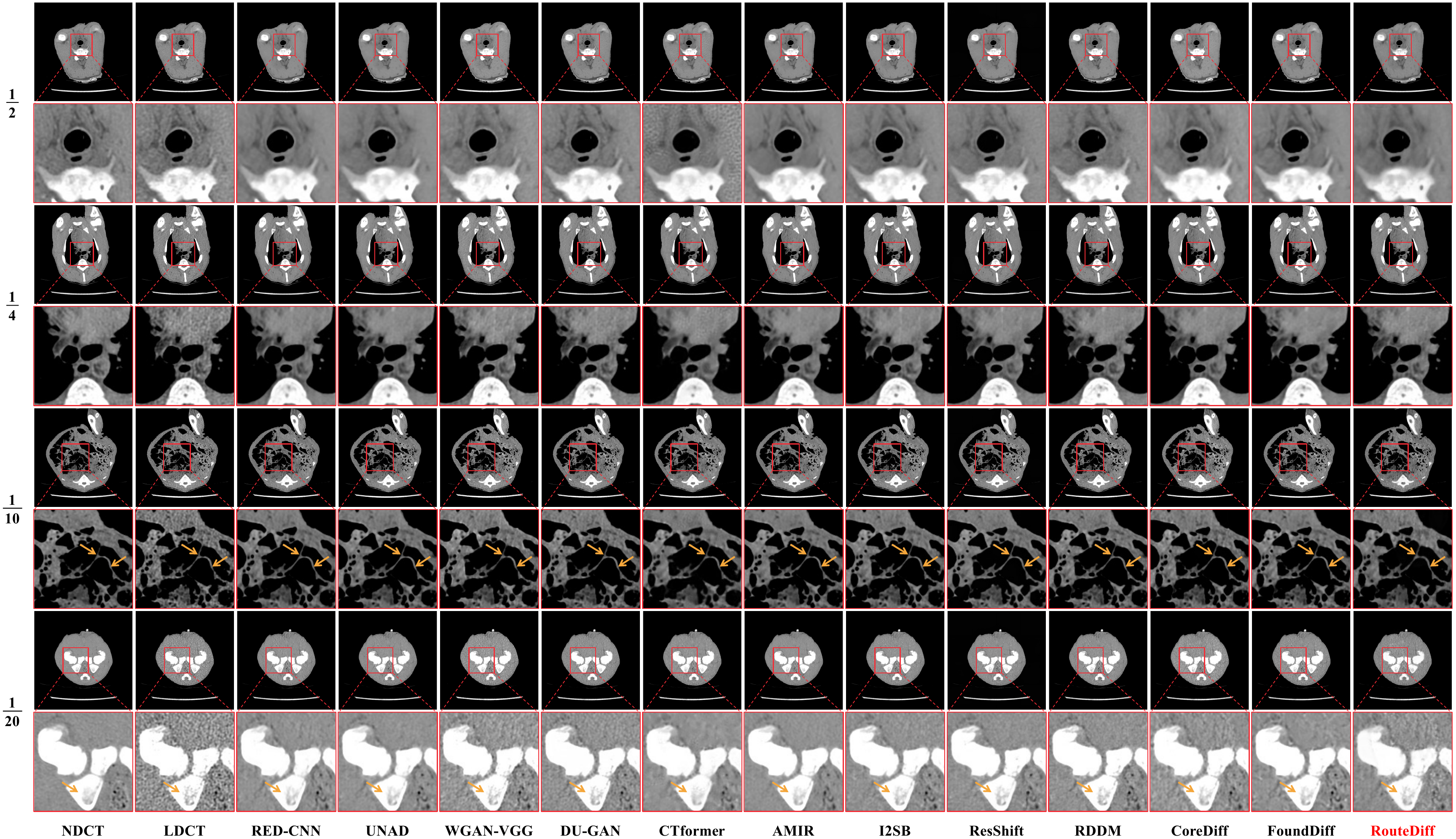}
\caption{Visual results of different methods on the Piglet dataset at $1/2$, $1/4$, $1/10$ and $1/20$ dose (top to bottom).}
\label{fig:piglet}
\end{figure*}

\begin{table*}[t]
\centering
\caption{Comparison results of different methods on the Piglet dataset with models trained on Mayo 2020 (mean $\pm$ standard deviation). Seen and unseen denote doses included in and absent from the training data. The best and second-best results are highlighted in bold and underlined, respectively.}
\label{tab:piglet}
\footnotesize
\setlength{\tabcolsep}{2pt}
\begin{tabular*}{\textwidth}{@{\extracolsep{\fill}}l*{8}{c}}
\toprule
\multirow{2}{*}{Methods} & \multicolumn{2}{c}{$1/2$ (150~mAs), unseen} & \multicolumn{2}{c}{$1/4$ (75~mAs), seen} & \multicolumn{2}{c}{$1/10$ (30~mAs), seen} & \multicolumn{2}{c}{$1/20$ (15~mAs), unseen}\\
\cmidrule(lr){2-3}\cmidrule(lr){4-5}\cmidrule(lr){6-7}\cmidrule(lr){8-9}
 & PSNR (dB)$\uparrow$ & SSIM (\%)$\uparrow$ & PSNR (dB)$\uparrow$ & SSIM (\%)$\uparrow$ & PSNR (dB)$\uparrow$ & SSIM (\%)$\uparrow$ & PSNR (dB)$\uparrow$ & SSIM (\%)$\uparrow$\\
\midrule
RED-CNN & 35.55$\pm$0.59 & 96.19$\pm$0.25 & 34.78$\pm$0.64 & 94.72$\pm$0.32 & 34.32$\pm$1.21 & 93.75$\pm$0.23 & 31.50$\pm$1.89 & 88.35$\pm$0.64\\
UNAD & 35.82$\pm$0.71 & 96.32$\pm$0.35 & 35.03$\pm$0.59 & 94.87$\pm$0.25 & 34.21$\pm$1.34 & 93.90$\pm$0.56 & 31.66$\pm$1.73 & 88.93$\pm$0.56\\
WGAN-VGG & 36.10$\pm$0.75 & 96.43$\pm$0.37 & 34.25$\pm$0.21 & 94.42$\pm$0.32 & 33.87$\pm$1.42 & 93.43$\pm$0.64 & 30.25$\pm$1.87 & 86.85$\pm$0.43\\
DU-GAN & 36.46$\pm$0.78 & 96.46$\pm$0.18 & 35.13$\pm$0.32 & 94.25$\pm$0.17 & 34.14$\pm$1.62 & 93.86$\pm$0.53 & 29.82$\pm$1.72 & 86.81$\pm$0.56\\
CTformer & 36.13$\pm$0.76 & 95.50$\pm$0.25 & 34.43$\pm$0.64 & 94.17$\pm$0.19 & 33.89$\pm$1.18 & 92.38$\pm$0.53 & 30.39$\pm$1.45 & 86.76$\pm$0.64\\
AMIR & 36.25$\pm$0.52 & 95.89$\pm$0.43 & 35.24$\pm$0.32 & 93.84$\pm$0.23 & 34.09$\pm$1.12 & 93.29$\pm$0.64 & 31.70$\pm$1.82 & 86.79$\pm$0.76\\
I2SB & \underline{36.66$\pm$0.47} & \underline{96.82$\pm$0.21} & 34.97$\pm$0.31 & 94.67$\pm$0.14 & 34.36$\pm$1.45 & \underline{94.26$\pm$0.72} & 31.37$\pm$1.72 & 88.16$\pm$0.23\\
ResShift & 34.94$\pm$0.42 & 96.24$\pm$0.36 & 34.38$\pm$0.18 & 94.79$\pm$0.12 & 33.71$\pm$1.52 & 93.83$\pm$0.14 & 28.92$\pm$1.62 & 87.10$\pm$0.41\\
RDDM & 36.59$\pm$0.53 & 96.26$\pm$0.26 & 34.83$\pm$0.14 & 94.85$\pm$0.17 & 33.82$\pm$1.09 & 93.86$\pm$0.32 & 30.21$\pm$1.54 & 86.99$\pm$0.42\\
CoreDiff & 36.40$\pm$0.43 & 96.32$\pm$0.31 & 35.27$\pm$0.32 & \underline{95.17$\pm$0.27} & 34.17$\pm$1.32 & 93.98$\pm$0.42 & 31.32$\pm$1.52 & 89.04$\pm$0.33\\
FoundDiff & 36.54$\pm$0.67 & 96.35$\pm$0.21 & \underline{35.32$\pm$0.21} & 95.00$\pm$0.21 & \underline{34.39$\pm$1.23} & 94.02$\pm$0.41 & \underline{31.78$\pm$1.78} & \underline{89.20$\pm$0.58}\\
\midrule
RouteDiff (ours) & \textbf{37.21$\pm$0.73} & \textbf{97.12$\pm$0.19} & \textbf{35.58$\pm$0.18} & \textbf{95.21$\pm$0.23} & \textbf{34.71$\pm$1.42} & \textbf{94.53$\pm$0.32} & \textbf{32.45$\pm$1.46} & \textbf{89.67$\pm$0.28}\\
\bottomrule
\end{tabular*}
\end{table*}

\subsection{Comparison results on the Mayo 2020 dataset}
\label{sec:res_mayo20}

On Mayo 2020, a single model restores chest and abdominal scans at both $1/4$ and $1/10$ dose. As shown in Table~\ref{tab:mayo20}, RouteDiff achieves the highest PSNR (40.95~dB) and SSIM (95.13\%), together with the highest VIF. Its PSNR and SSIM exceed those of the respective runners-up, FoundDiff and AMIR, by 0.61~dB and 0.31\%, and its PSNR is 0.69~dB above CoreDiff.

Fig.~\ref{fig:mayo20} presents the visual results. WGAN-VGG leaves residual noise in the liver (a), and CTformer produces patchy, artifact-like texture. RED-CNN and, most visibly, FoundDiff over-smooth: the hepatic parenchyma around the vessels marked by the arrows in (a) loses its fine-grained texture, and FoundDiff smooths the renal medulla at the arrow in (b) into a featureless gray area. In the $1/10$-dose chest cases (c) and (d), where the input noise is strongest, RED-CNN, AMIR and FoundDiff also blur the trabecular pattern of the vertebral bodies at the arrows. In contrast, RouteDiff keeps the parenchymal texture and the renal corticomedullary pattern close to the routine-dose images and retains clearly more trabecular texture than these methods in (c) and (d). Across both anatomies and both doses, RouteDiff thus suppresses noise while preserving fine anatomical structure better than the smoothing baselines.

\begin{figure*}[t]
\centering
\includegraphics[width=\textwidth]{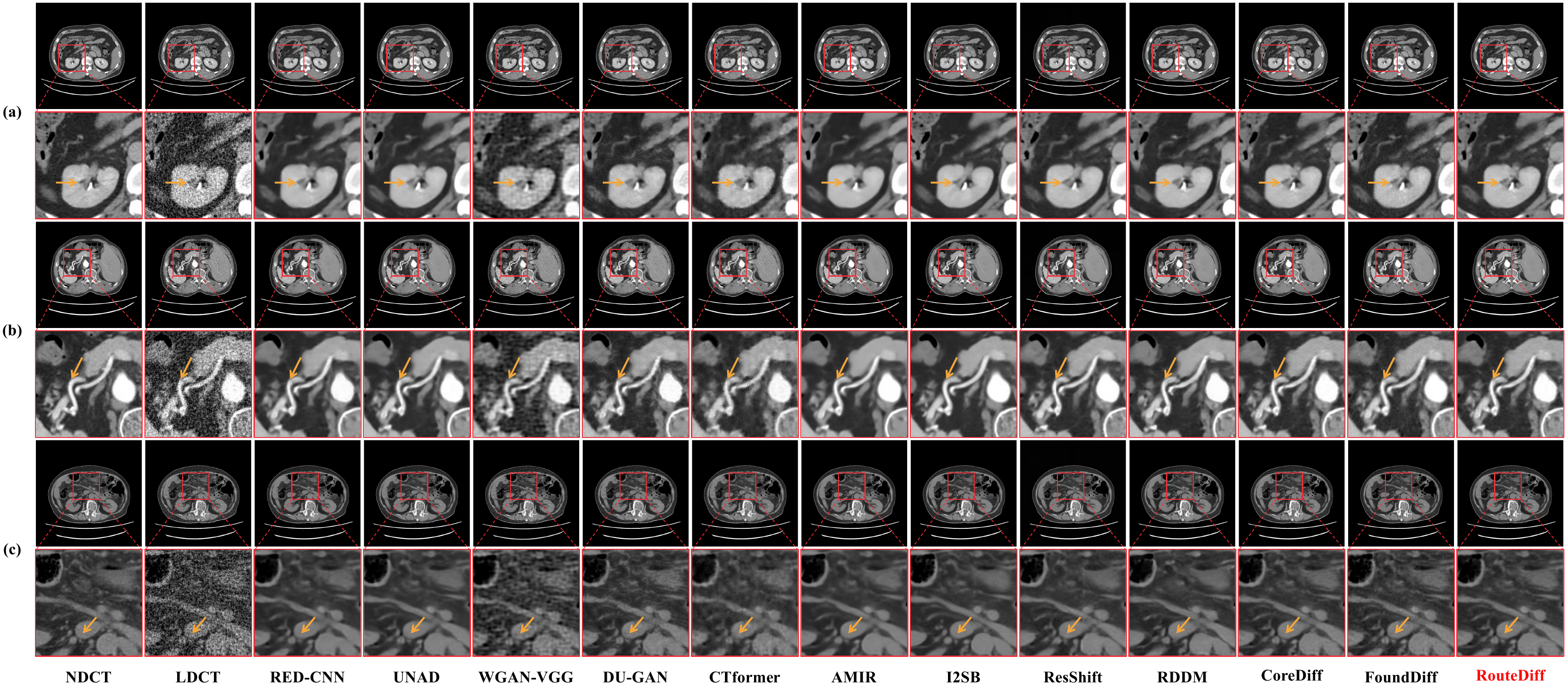}
\caption{Visual results of different methods on the AMOS dataset.}
\label{fig:amos}
\end{figure*}

\begin{table*}[tp]
\centering
\caption{Restoration and organ segmentation results of different methods on the AMOS dataset. Dice (\%) and 95th-percentile Hausdorff distance (HD95, mm) of the frozen nnU-Net \cite{isensee2021nnunet} are averaged over the 15 classes. Spl: spleen; RK: right kidney; LK: left kidney; GB: gallbladder; Eso: esophagus; Liv: liver; Sto: stomach; Ao: aorta; IVC: inferior vena cava; Pan: pancreas; RAG: right adrenal gland; LAG: left adrenal gland; Duo: duodenum; Bla: bladder; P/U: prostate/uterus. NDCT (routine-dose reference) is not ranked. The best and second-best results are highlighted in bold and underlined, respectively.}
\label{tab:amos_seg}
\scriptsize
\setlength{\tabcolsep}{1.5pt}
\begin{tabular*}{\textwidth}{@{\extracolsep{\fill}}l*{19}{c}}
\toprule
\multirow{2}{*}{Methods} & \multicolumn{2}{c}{Restoration} & \multicolumn{2}{c}{Segmentation} & \multicolumn{15}{c}{Organ-wise Dice (\%)}\\
\cmidrule(lr){2-3}\cmidrule(lr){4-5}\cmidrule(lr){6-20}
 & PSNR (dB)$\uparrow$ & SSIM (\%)$\uparrow$ & Dice (\%)$\uparrow$ & HD95 (mm)$\downarrow$ & Spl & RK & LK & GB & Eso & Liv & Sto & Ao & IVC & Pan & RAG & LAG & Duo & Bla & P/U\\
\midrule
RED-CNN & 36.72 & 89.78 & 79.62 & 49.01 & 86.79 & 89.38 & 88.39 & 70.57 & 77.56 & 92.78 & 89.77 & \underline{88.73} & 81.16 & 79.97 & 60.13 & 60.42 & 72.72 & 79.45 & 76.41\\
UNAD & 36.63 & 90.97 & 78.64 & 22.70 & 86.40 & 90.06 & 88.74 & 68.84 & 76.30 & 93.62 & 90.19 & 86.07 & 81.27 & 78.41 & 57.15 & 58.28 & 72.39 & 79.11 & 72.77\\
WGAN-VGG & 34.58 & 88.94 & 76.19 & 64.63 & 86.01 & 85.31 & 86.33 & 63.75 & 73.29 & 92.03 & 87.89 & 87.02 & 81.18 & 77.50 & 50.87 & 54.22 & 69.79 & 76.45 & 71.16\\
DU-GAN & 36.13 & 88.86 & 77.86 & 27.39 & 86.57 & 88.03 & 87.89 & 67.92 & 75.87 & 92.73 & 86.22 & 84.60 & 81.07 & 78.50 & 57.43 & 58.47 & 70.02 & 79.51 & 73.02\\
CTformer & 36.61 & 87.33 & 77.56 & 53.60 & 86.02 & 89.83 & 88.97 & 66.70 & 77.53 & 93.55 & 89.57 & 83.33 & 80.22 & 80.14 & 53.91 & 57.06 & 73.25 & 77.87 & 65.50\\
AMIR & 36.78 & 89.98 & 78.52 & 37.03 & 86.61 & 89.70 & 88.38 & 68.85 & 77.21 & 92.72 & 88.64 & 84.74 & 81.23 & 79.93 & 58.19 & 58.12 & 72.14 & 79.78 & 71.51\\
I2SB & 36.57 & \underline{91.83} & 79.56 & \underline{14.11} & 86.28 & 89.12 & 88.97 & 73.37 & 76.11 & \underline{94.33} & 85.07 & 87.82 & 81.10 & 78.66 & 60.10 & 61.43 & 70.27 & 79.48 & 81.36\\
ResShift & 36.68 & 91.22 & 79.89 & 21.02 & 85.89 & 89.83 & 88.97 & 72.47 & 76.35 & 93.39 & 89.95 & 88.33 & 81.27 & 78.24 & 60.46 & 59.85 & \textbf{75.00} & 78.86 & 79.49\\
RDDM & 37.26 & 90.92 & \underline{80.30} & 21.43 & \underline{87.19} & \underline{90.28} & \underline{89.01} & 73.52 & \underline{78.06} & \textbf{94.97} & \underline{90.38} & 82.63 & \underline{81.31} & 80.01 & \underline{60.73} & \textbf{62.77} & \underline{73.63} & \underline{79.94} & 80.02\\
CoreDiff & 37.31 & 91.17 & 80.18 & 16.04 & 85.84 & 90.03 & \textbf{90.50} & 72.12 & 77.85 & 93.73 & 89.30 & \textbf{89.24} & \textbf{81.50} & 78.90 & 59.47 & 60.20 & 73.11 & 79.51 & \underline{81.45}\\
FoundDiff & \underline{37.33} & 91.58 & 80.15 & 19.41 & 86.79 & \textbf{90.42} & 88.57 & \underline{74.07} & 77.61 & 93.78 & 90.10 & 85.33 & 81.24 & \underline{80.24} & 60.57 & \underline{61.46} & 72.63 & 79.74 & 79.70\\
\midrule
RouteDiff (ours) & \textbf{37.79} & \textbf{92.63} & \textbf{80.92} & \textbf{13.92} & \textbf{87.90} & 89.92 & 88.97 & \textbf{75.20} & \textbf{79.30} & 92.78 & \textbf{91.13} & 87.48 & 81.29 & \textbf{81.20} & \textbf{64.30} & 60.69 & 70.35 & \textbf{81.23} & \textbf{82.10}\\
\midrule
NDCT & -- & -- & 88.46 & 12.99 & 96.05 & 95.19 & 95.15 & 77.92 & 85.35 & 96.66 & 92.45 & 95.63 & 88.83 & 85.65 & 76.96 & 77.96 & 82.02 & 91.79 & 89.27\\
\bottomrule
\end{tabular*}
\end{table*}

\subsection{Comparison results on the Mayo 2016 dataset}
\label{sec:res_mayo16}

As shown in Table~\ref{tab:mayo16}, RouteDiff achieves the best values on all five metrics on Mayo 2016, with a PSNR of 45.18~dB and an SSIM of 97.23\%, which exceed those of the respective runners-up, CoreDiff and FoundDiff, by 1.02~dB and 0.49\%.

In Fig.~\ref{fig:mayo16}, WGAN-VGG retains residual noise in the peripancreatic (a) and mesenteric (b) soft tissue, and CTformer leaves patchy texture (b). The smoothing methods suppress the noise at the cost of structure: RED-CNN and FoundDiff smooth the pancreatic margin in (a), and RED-CNN, AMIR and FoundDiff blur the trabecular bone of the pelvis in (c), both marked by the arrows. RouteDiff preserves the margin and the trabecular pattern and retains a fine-grained texture closer to that of the routine-dose images. Mayo 2016 thus confirms that the higher fidelity of RouteDiff is not obtained by over-smoothing.

\subsection{Comparison results on the Piglet dataset}
\label{sec:res_piglet}

Although no Piglet image is used in training, RouteDiff achieves the highest PSNR and SSIM in Table~\ref{tab:piglet} at all four doses, including the two unseen ones. Over CoreDiff, its PSNR gain is larger at the unseen doses (0.81 and 1.13~dB) than at the seen doses (0.31 and 0.54~dB). Its smallest margin over the runner-up at an unseen dose (0.55~dB over I2SB at $1/2$ dose) still exceeds the largest at a seen dose (0.32~dB).

In Fig.~\ref{fig:piglet}, the real noise is fine-grained rather than streak-like and coarsens from $1/2$ dose (top) to $1/20$ dose (bottom). CTformer leaves a periodic texture at $1/2$ dose. At $1/10$ dose, RouteDiff preserves the thin walls between the gas-filled bowel loops marked by the arrows. At $1/20$ dose, RouteDiff keeps the bone outline and marrow pattern at the arrow, whereas RED-CNN and AMIR, whose backgrounds are smoother, still have a lower PSNR at this dose. RouteDiff thus transfers to real scanner noise and gains most over its baseline at the doses absent from training, which highlights its robustness to unseen acquisition conditions.

\subsection{Comparison results on the AMOS dataset}
\label{sec:res_amos}

As shown in the restoration columns of Table~\ref{tab:amos_seg}, RouteDiff achieves the highest PSNR (37.79~dB) and SSIM (92.63\%) on AMOS, 0.46~dB above FoundDiff and 0.80\% above I2SB, respectively, and exceeds CoreDiff by 0.48~dB and 1.46\%.

In Fig.~\ref{fig:amos}, RED-CNN, UNAD and AMIR flatten the soft tissue into a plastic-like appearance. WGAN-VGG, CTformer and FoundDiff instead leave residual noise or blotchy texture, WGAN-VGG adds a grid pattern (b), and I2SB and ResShift show straight tile seams, clearly in (c) and faintly in (a). RouteDiff gives clean organ interiors without such artifacts while keeping the renal hilum (a), the tortuous artery (b) and the vessel outline (c) marked by the arrows close to the routine-dose images. With low-dose inputs reconstructed by iterative OS-SART, RouteDiff thus again combines noise suppression with structural fidelity.

\begin{figure*}[t]
\centering
\includegraphics[width=\textwidth]{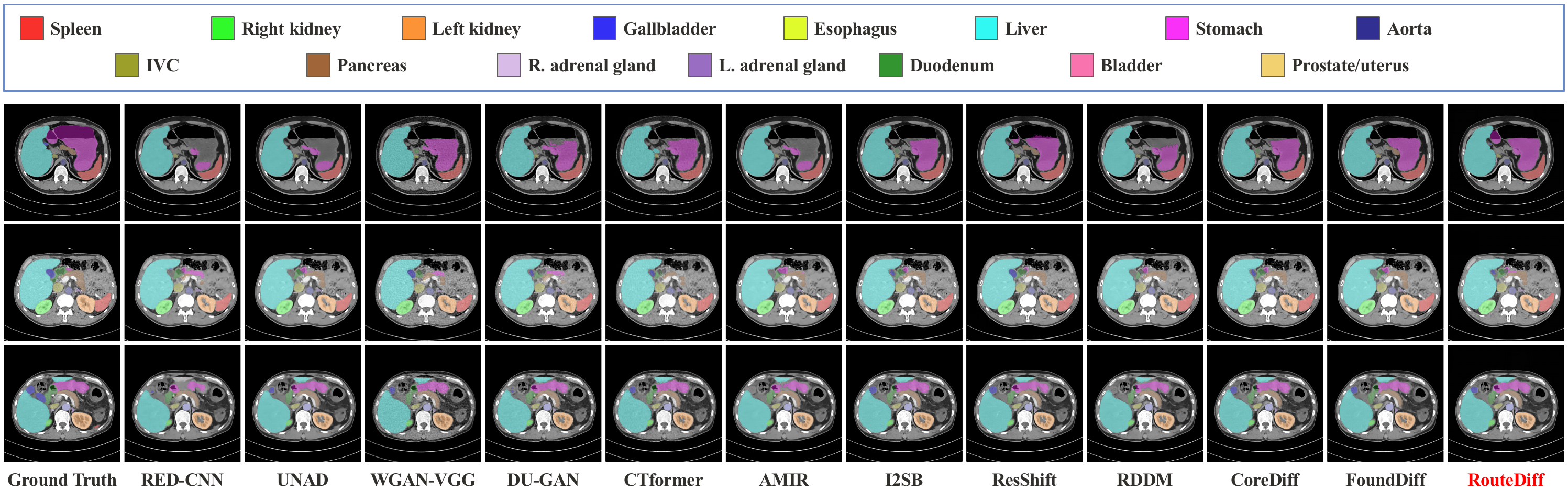}
\caption{Visual segmentation results of the frozen nnU-Net on the AMOS images restored by different methods.}
\label{fig:amos_seg}
\end{figure*}

To examine whether the restored images benefit downstream analysis, we train nnU-Net \cite{isensee2021nnunet} on the routine-dose AMOS training images, freeze it and apply it to the restored test slices. We report the Dice score and the 95th-percentile Hausdorff distance (HD95), with the routine-dose test images (NDCT) as the upper reference. As shown in Table~\ref{tab:amos_seg}, segmentation on the outputs of RouteDiff yields the highest Dice (80.92\%) and the lowest HD95 (13.92~mm) among the restoration methods. RouteDiff obtains the best organ-wise Dice on eight of the 15 classes, including small or low-contrast organs whose boundaries are easily obscured by noise, namely the right adrenal gland, the gallbladder, the esophagus and the pancreas. Its largest margin over the runner-up (3.57 points) is on the right adrenal gland. In the first slice of Fig.~\ref{fig:amos_seg}, RED-CNN, UNAD and AMIR leave most of the stomach unlabeled, whereas only the segmentation of the RouteDiff output covers the small anterior part of the stomach in full, as in the ground truth, consistent with its highest stomach Dice. By mean Dice, the outputs of RouteDiff are thus the most useful to a network that has seen only routine-dose data, so the restoration benefits downstream analysis beyond pixel-level fidelity.

\subsection{Computational cost}
\label{sec:cost}

\begin{table}[t]
\centering
\caption{Computational cost of different methods on one RTX~5090 GPU. The inference time is measured per $512\times512$ slice.}
\label{tab:cost}
\footnotesize
\setlength{\tabcolsep}{3pt}
\begin{tabular*}{\columnwidth}{@{\extracolsep{\fill}}lcccc}
\toprule
Methods & Params (M) & Batch size & Training (h/epoch) & Inference (s)\\
\midrule
RED-CNN & 1.83 & 4 & 0.84 & 0.034\\
AMIR & 12.72 & 4 & 2.23 & 0.132\\
CoreDiff & 6.51 & 4 & 1.32 & 0.159\\
RDDM & 18.94 & 2 & 3.43 & 0.242\\
FoundDiff & 21.64 & 2 & 3.96 & 0.342\\
\midrule
RouteDiff (ours) & 7.23 & 4 & 3.07 & 0.183\\
\bottomrule
\end{tabular*}
\end{table}

Table~\ref{tab:cost} compares the cost of RouteDiff with that of representative convolutional neural network (CNN), transformer and diffusion-based methods. RouteDiff has 7.23~M parameters; the 0.72~M added to CoreDiff come from the second prediction head and HSR, and EAR adds none. With the same ten NFE as CoreDiff, RouteDiff needs 0.183~s per slice, only 0.024~s more than CoreDiff and less than RDDM and FoundDiff. An epoch takes 2.3 times as long as for CoreDiff because each iteration executes two or three calls (Section~\ref{sec:training}), yet still less than for RDDM and FoundDiff. Among the diffusion-based methods, RouteDiff thus offers a favorable trade-off between efficiency and restoration quality.

\subsection{Ablation study}
\label{sec:ablation}

\subsubsection{Contribution of each module}

\begin{table}[t]
\centering
\caption{Ablation study of the three modules on the Mayo 2016 dataset (mean $\pm$ standard deviation). Setting~1 is the CoreDiff baseline. The best and second-best results are highlighted in bold and underlined, respectively.}
\label{tab:ablation}
\footnotesize
\setlength{\tabcolsep}{2.5pt}
\begin{tabular*}{\columnwidth}{@{\extracolsep{\fill}}cccccc}
\toprule
Setting & FRR & HSR & EAR & PSNR (dB)$\uparrow$ & SSIM (\%)$\uparrow$ \\
\midrule
1 & \xmark & \xmark & \xmark & 44.16$\pm$0.26 & 96.36$\pm$0.10 \\
2 & \cmark & \xmark & \xmark & 44.43$\pm$0.26 & 96.85$\pm$0.08 \\
3 & \cmark & \cmark & \xmark & 44.72$\pm$0.25 & 97.12$\pm$0.07 \\
4 & \cmark & \xmark & \cmark & \underline{44.99$\pm$0.25} & \underline{97.18$\pm$0.07} \\
5 & \cmark & \cmark & \cmark & \textbf{45.18$\pm$0.25} & \textbf{97.23$\pm$0.06} \\
\bottomrule
\end{tabular*}
\end{table}

Setting~1 is the CoreDiff baseline, a U-Net with a single prediction head, timestep conditioning and the regular update of Eq.~\eqref{eq:update}. Settings~2--5 add the modules marked in Table~\ref{tab:ablation} and share the backbone, initialization, iterations, branch sampling and NFE per training iteration. In the settings without EAR, the transition from step 3 to step 2 is a regular update.

As shown in Table~\ref{tab:ablation}, every added module improves both metrics, and the full model is the best on both. FRR improves PSNR by 0.27~dB over Setting~1, which already places Setting~2 above every competing method of Table~\ref{tab:mayo16} in both PSNR and SSIM. Adding HSR, including its FiLM and FAFE modulation (Setting~3), improves PSNR by a further 0.29~dB, whereas adding EAR instead (Setting~4) improves it by 0.56~dB over Setting~2. The two modules are complementary: each still improves both metrics when added to the other, with PSNR gains of 0.19~dB for HSR (Setting~5 vs.\ Setting~4) and 0.46~dB for EAR (Setting~5 vs.\ Setting~3). This complementarity arises because HSR provides better routes for the anchor and EAR hands the re-entered state back to HSR to be judged again (Table~\ref{tab:design}(c)). Together, the three modules add 1.02~dB in PSNR over the baseline for 0.72~M parameters and no additional NFE, of which 0.75~dB comes from adding HSR and EAR on top of FRR.

\subsubsection{Design analysis}
\begin{table}[t]
\centering
\caption{Design analysis on the Mayo 2020 dataset. Each variant changes one design choice of the full model. The best and second-best results are highlighted in bold and underlined, respectively.}
\label{tab:design}
\footnotesize
\setlength{\tabcolsep}{3pt}
\begin{tabular*}{\columnwidth}{@{\extracolsep{\fill}}lcc}
\toprule
Variant & PSNR (dB)$\uparrow$ & SSIM (\%)$\uparrow$\\
\midrule
Full model (RouteDiff) & \textbf{40.95} & \textbf{95.13}\\
\midrule
\multicolumn{3}{@{}l}{\textit{(a) FRR: where detail is admitted}}\\
\quad Fidelity only ($c=0$) & 40.60 & 94.87\\
\quad Refinement only ($c=1$) & 40.48 & 94.55\\
\quad Global trade-off ($\bar{c}$) & 40.78 & 95.04\\
\midrule
\multicolumn{3}{@{}l}{\textit{(b) HSR: stage judgment and grouping}}\\
\quad Timestep-only $q_k$ & 40.70 & 94.84\\
\quad w/o route evidence & \underline{40.84} & 94.99\\
\quad Hypergraph $\to$ convolution & 40.73 & 94.90\\
\quad Hypergraph $\to$ self-attention & 40.76 & 95.04\\
\quad Hypergraph $\to$ $K$-NN graph & 40.80 & 95.07\\
\quad w/o skip interaction & 40.83 & 95.08\\
\midrule
\multicolumn{3}{@{}l}{\textit{(c) EAR: what is re-entered}}\\
\quad Anchor $\hat{y}_m$ & 40.74 & 95.05\\
\quad w/o re-projection & 40.56 & 94.80\\
\quad Keep $q_k$ at re-entry & 40.82 & 95.06\\
\quad Clear $\mathcal{H}_m$ & \underline{40.84} & \underline{95.09}\\
\bottomrule
\end{tabular*}
\end{table}

Table~\ref{tab:design} groups the design choices by the three questions raised in Section~\ref{sec:intro}. Each variant is retrained with the same protocol on Mayo 2020, whose test set pools both doses and anatomies. All 13 variants fall below the full model in both PSNR and SSIM, by 0.11 to 0.47~dB in PSNR, confirming that every examined design choice contributes to the final performance.

\textit{Where detail is admitted.} The routes are fused with $c=0$, with $c=1$ or with the per-image mean $\bar{c}$ of the confidence map, recomputed at every call, while the route loss and the evidence keep using $c$; the same $c=0$, $c=1$ or $\bar{c}$ is used in Eq.~\eqref{eq:xest}. Taking all high frequencies from the fidelity route discards the trusted detail of the refinement route, and taking them all from the refinement route also admits its untrusted texture; the two variants lower PSNR by 0.35~dB and 0.47~dB, respectively. The global trade-off $\bar{c}$ admits detail with the same mean weight as the confidence map but still falls 0.17~dB in PSNR behind the full model, so the gain comes from where detail is admitted rather than from how much.

\textit{Stage judgment and grouping.} For the timestep-only variant, the state signature is set to zero in $f_{init}$ and $f_{upd}$ of Eq.~\eqref{eq:progress}, while the ordered update, the prior term, the hypergraph and the skip interaction are kept; for the variant without route evidence, $\bar{\varepsilon}_{k-1}=0$ in Eq.~\eqref{eq:node} at every call. They lower PSNR by 0.25~dB and 0.11~dB, respectively; the timestep-only variant shows the largest drop in group (b), so the stage is better judged from the observed state than from the timestep alone, and the route evidence of FRR contributes to this judgment. Each relation operator replaces the hypergraph in both the global readout and the skip interaction. The node states are mapped by a stack of $3\times3$ convolutions, by global multi-head self-attention at the bottleneck (windowed at the skip levels) or by graph convolution over the $K=8$ nearest positions in the state space, as in Vision GNN \cite{han2022vig}. In all three cases, the readout is obtained by mean and max pooling. With parameters and GFLOPs matched to those of the hypergraph module within 5\%, the three replacements lower PSNR by 0.22, 0.19 and 0.15~dB, so the advantage of the hypergraph is not explained by additional capacity; removing only the skip interaction lowers PSNR by 0.12~dB. The drop shrinks from local convolution to the $K$-NN graph, which relates positions by their state rather than by their location. Yet none of these operators reaches the full model, which shows that grouping scattered regions in the same state is more effective than local or pairwise aggregation.

\begin{figure}[t]
\centering
\includegraphics[width=\columnwidth]{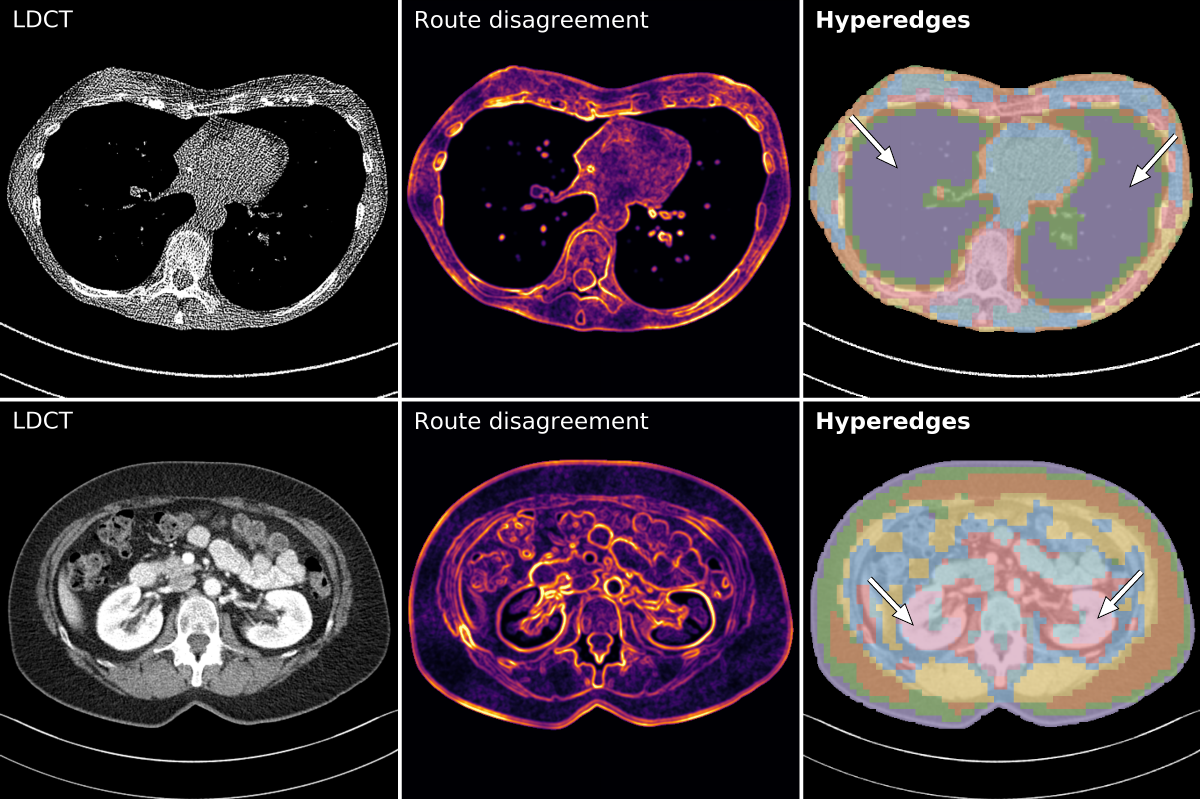}
\caption{Visualization of the state hypergraph at step $m$ on the Mayo 2020 dataset for a chest slice (top) and an abdominal slice (bottom). From left to right: low-dose input, route disagreement $|\Delta^{hf}|$ and dominant hyperedge $\arg\max_g A_{pg}$ of each bottleneck position.}
\label{fig:mechanism}
\end{figure}

\textit{What is re-entered.} Using the fused prediction $\hat{y}_m$ as the anchor, without the context-based alignment and clipping of Eq.~\eqref{eq:estparts}, lowers PSNR by 0.21~dB, and skipping the re-projection ($x_r=\mathrm{sg}(x_{est})$) lowers it by 0.39~dB, the largest drop in group (c) and the second largest in Table~\ref{tab:design}. At re-entry, keeping the previous stage judgment ($\mathcal{K}_0$ contains only the first call) lowers PSNR by 0.13~dB, and clearing the evidence ($\mathcal{H}_m=\mathbf{0}$ and $v_k=0$ at the re-entered call) lowers it by 0.11~dB. Re-entry is therefore most effective when a route-aware anchor is projected back onto the trajectory; the re-entered state has to be judged again, and the evidence observed before re-entry remains informative for this judgment.

\subsubsection{Hypergraph visualization and sensitivity}
\label{sec:hyper}

Fig.~\ref{fig:mechanism} visualizes the state hypergraph of the full model at step $m$. In the chest slice, both lung regions (arrows) are dominated by the same hyperedge although they are separated by the mediastinum, and in the abdominal slice, both kidneys (arrows) fall into one hyperedge, which also covers the vertebral arch behind the vertebral body. The boundaries on which the routes disagree, visible as bright contours in the $|\Delta^{hf}|$ maps, are assigned to hyperedges other than those of the adjacent homogeneous tissue and form bands along the body outline and around the kidneys. The hyperedges thus group distant positions that share similar features and route evidence rather than spatially adjacent ones, a relation that the local and pairwise operators of Table~\ref{tab:design}(b) do not represent explicitly and that is consistent with their lower PSNR.

\begin{figure}[t]
\centering
\includegraphics[width=\columnwidth]{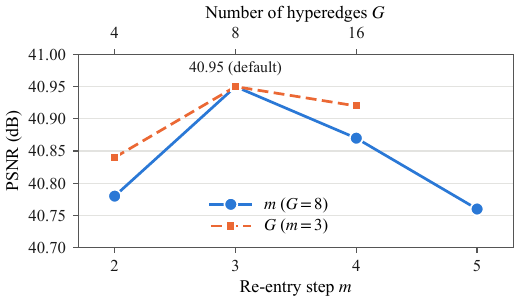}
\caption{Sensitivity analysis on the Mayo 2020 dataset. Solid blue curve: the re-entry step $m$ is varied (bottom axis) with $G=8$ fixed. Dashed orange curve: the number of hyperedges $G$ is varied (top axis) with $m=3$ fixed. The two sweeps are independent and share only the default setting ($m=3$, $G=8$).}
\label{fig:sensitivity}
\end{figure}

We vary one of the two structural hyperparameters at a time while the other is kept at its default value, the re-entry step $m\in\{2,3,4,5\}$ with $G=8$ and the number of hyperedges $G\in\{4,8,16\}$ with $m=3$, and retrain the model for each value as in Section~\ref{sec:training}. As shown in Fig.~\ref{fig:sensitivity}, with the other parameter fixed, $m=3$ and $G=8$ each yield the highest PSNR on Mayo 2020, and the PSNR varies by at most 0.19~dB over the tested ranges. We attribute the drop at larger $m$ to an anchor built from less reliable routes and the drop at smaller $m$ to fewer calls being left to refine the re-entered state. Too few hyperedges cost more than too many (0.11~dB for $G=4$ against 0.03~dB for $G=16$), as they merge regions in different states, whereas too many fragment regions in the same state. RouteDiff is thus robust to $m$ and $G$ within the tested ranges.

\let\FloatBarrier\relax
\section{Conclusion}
\label{sec:conclusion}
Existing diffusion samplers for LDCT restoration apply one structure--detail trade-off to the whole image, set the restoration emphasis of each step by the timestep and derive every next state recursively from the current one. To address these problems, this paper proposes RouteDiff, a state-aware dual-route diffusion framework. FRR predicts a fidelity route and a refinement route with one shared U-Net, admits refinement detail pixel-wise where a confidence map trusts it, and exposes the route disagreement as evidence. On this basis, HSR organizes the bottleneck features and this evidence into a state hypergraph, whose readout sets an ordered stage judgment for the bottleneck and whose incidence guides feature interaction in the skip connections. EAR then re-projects a route-aware anchor onto the trajectory at a fixed step and hands the re-entered state back to HSR to be judged again. Experiments on four CT datasets show that RouteDiff, at the same NFE as its baseline, achieves the best PSNR and SSIM on all four datasets, generalizes to the unseen doses of real scans and yields the highest Dice among the compared restoration methods in downstream organ segmentation.

Although RouteDiff performs well across datasets with different anatomies, doses and noise sources, several limitations remain. First, the re-entry step is fixed for all slices rather than chosen from the observed state; selecting it adaptively could further exploit the stage judgment of HSR. Second, RouteDiff restores each slice in two dimensions with a three-slice context, so extending the state hypergraph to volumetric restoration and assessing the restored images in a reader study by radiologists are worth investigating.

\section*{Declaration of competing interest}
The authors declare that they have no known competing financial interests or personal relationships that could have appeared to influence the work reported in this paper.

\section*{Data availability}
The datasets used in this study (Mayo 2020, Mayo 2016, Piglet and AMOS) are available from the sources cited in Section~\ref{sec:datasets}. The code will be made publicly available.

\section*{Declaration of generative AI and AI-assisted technologies in the manuscript preparation process}
During the preparation of this work the authors used Claude (Anthropic) in order to improve the language and readability of the manuscript. After using this tool, the authors reviewed and edited the content as needed and take full responsibility for the content of the published article.

\bibliographystyle{elsarticle-num}
\bibliography{refs}

\begin{thebibliography}{10}
\expandafter\ifx\csname url\endcsname\relax
  \def\url#1{\texttt{#1}}\fi
\expandafter\ifx\csname urlprefix\endcsname\relax\def\urlprefix{URL }\fi
\expandafter\ifx\csname href\endcsname\relax
  \def\href#1#2{#2} \def\path#1{#1}\fi

\bibitem{smith2009radiation}
R.~Smith-Bindman, J.~Lipson, R.~Marcus, K.-P. Kim, M.~Mahesh, R.~Gould,
  A.~{Berrington de Gonz{\'a}lez}, D.~L. Miglioretti, Radiation dose associated
  with common computed tomography examinations and the associated lifetime
  attributable risk of cancer, Archives of Internal Medicine 169~(22) (2009)
  2078--2086.

\bibitem{redcnn}
H.~Chen, Y.~Zhang, M.~K. Kalra, F.~Lin, Y.~Chen, P.~Liao, J.~Zhou, G.~Wang,
  Low-dose {CT} with a residual encoder-decoder convolutional neural network,
  IEEE Transactions on Medical Imaging 36~(12) (2017) 2524--2535.

\bibitem{dugan}
Z.~Huang, J.~Zhang, Y.~Zhang, H.~Shan, {DU-GAN}: Generative adversarial
  networks with dual-domain {U-Net}-based discriminators for low-dose {CT}
  denoising, IEEE Transactions on Instrumentation and Measurement 71 (2022)
  1--12.

\bibitem{ctformer}
D.~Wang, F.~Fan, Z.~Wu, R.~Liu, F.~Wang, H.~Yu, {CTformer}: Convolution-free
  {Token2Token} dilated vision transformer for low-dose {CT} denoising, Physics
  in Medicine \& Biology 68~(6) (2023) 065012.

\bibitem{kim2019performance}
B.~Kim, M.~Han, H.~Shim, J.~Baek, A performance comparison of convolutional
  neural network-based image denoising methods: The effect of loss functions on
  low-dose {CT} images, Medical Physics 46~(9) (2019) 3906--3923.

\bibitem{ho2020denoising}
J.~Ho, A.~Jain, P.~Abbeel, Denoising diffusion probabilistic models, in:
  Advances in Neural Information Processing Systems, Vol.~33, 2020, pp.
  6840--6851.

\bibitem{song2020denoising}
J.~Song, C.~Meng, S.~Ermon, Denoising diffusion implicit models, in:
  International Conference on Learning Representations (ICLR), 2021.

\bibitem{i2sb}
G.-H. Liu, A.~Vahdat, D.-A. Huang, E.~A. Theodorou, W.~Nie, A.~Anandkumar,
  {I$^2$SB}: Image-to-image {Schr{\"o}dinger} bridge, in: International
  Conference on Machine Learning (ICML), Vol. 202 of PMLR, 2023, pp.
  22042--22062.

\bibitem{resshift}
Z.~Yue, J.~Wang, C.~C. Loy, {ResShift}: Efficient diffusion model for image
  super-resolution by residual shifting, in: Advances in Neural Information
  Processing Systems, Vol.~36, 2023, pp. 13294--13307.

\bibitem{rddm}
J.~Liu, Q.~Wang, H.~Fan, Y.~Wang, Y.~Tang, L.~Qu, Residual denoising diffusion
  models, in: Proceedings of the IEEE/CVF Conference on Computer Vision and
  Pattern Recognition (CVPR), 2024, pp. 2773--2783.

\bibitem{bansal2023cold}
A.~Bansal, E.~Borgnia, H.-M. Chu, J.~Li, H.~Kazemi, F.~Huang, M.~Goldblum,
  J.~Geiping, T.~Goldstein, Cold diffusion: Inverting arbitrary image
  transforms without noise, in: Advances in Neural Information Processing
  Systems, Vol.~36, 2023, pp. 41259--41282.

\bibitem{corediff}
Q.~Gao, Z.~Li, J.~Zhang, Y.~Zhang, H.~Shan, {CoreDiff}: Contextual
  error-modulated generalized diffusion model for low-dose {CT} denoising and
  generalization, IEEE Transactions on Medical Imaging 43~(2) (2024) 745--759.

\bibitem{founddiff}
Z.~Chen, Q.~Gao, Z.~Li, J.~Zhang, Y.~Zhang, J.~Zhao, H.~Shan, {FoundDiff}:
  Foundational diffusion model for generalizable low-dose {CT} denoising, IEEE
  Transactions on Medical Imaging 45~(8) (2026) 4366--4379.

\bibitem{bhadra2021hallucinations}
S.~Bhadra, V.~A. Kelkar, F.~J. Brooks, M.~A. Anastasio, On hallucinations in
  tomographic image reconstruction, IEEE Transactions on Medical Imaging
  40~(11) (2021) 3249--3260.

\bibitem{kalra2004techniques}
M.~K. Kalra, M.~M. Maher, T.~L. Toth, B.~Schmidt, B.~L. Westerman, H.~T.
  Morgan, S.~Saini, Techniques and applications of automatic tube current
  modulation for {CT}, Radiology 233~(3) (2004) 649--657.

\bibitem{blau2018perception}
Y.~Blau, T.~Michaeli, The perception-distortion tradeoff, in: Proceedings of
  the IEEE Conference on Computer Vision and Pattern Recognition (CVPR), 2018,
  pp. 6228--6237.

\bibitem{yi2018sharpness}
X.~Yi, P.~Babyn, Sharpness-aware low-dose {CT} denoising using conditional
  generative adversarial network, Journal of Digital Imaging 31~(5) (2018)
  655--669.

\bibitem{whang2022deblurring}
J.~Whang, M.~Delbracio, H.~Talebi, C.~Saharia, A.~G. Dimakis, P.~Milanfar,
  Deblurring via stochastic refinement, in: Proceedings of the IEEE/CVF
  Conference on Computer Vision and Pattern Recognition (CVPR), 2022, pp.
  16293--16303.

\bibitem{angelopoulos2022image}
A.~N. Angelopoulos, A.~P. Kohli, S.~Bates, M.~I. Jordan, J.~Malik, T.~Alshaabi,
  S.~Upadhyayula, Y.~Romano, Image-to-image regression with distribution-free
  uncertainty quantification and applications in imaging, in: International
  Conference on Machine Learning (ICML), Vol. 162 of PMLR, 2022, pp. 717--730.

\bibitem{ning2023input}
M.~Ning, E.~Sangineto, A.~Porrello, S.~Calderara, R.~Cucchiara, Input
  perturbation reduces exposure bias in diffusion models, in: International
  Conference on Machine Learning (ICML), Vol. 202 of PMLR, 2023, pp.
  26245--26265.

\bibitem{li2024alleviating}
M.~Li, T.~Qu, R.~Yao, W.~Sun, M.-F. Moens, Alleviating exposure bias in
  diffusion models through sampling with shifted time steps, in: International
  Conference on Learning Representations (ICLR), 2024.

\bibitem{chung2022come}
H.~Chung, B.~Sim, J.~C. Ye, {Come-Closer-Diffuse-Faster}: Accelerating
  conditional diffusion models for inverse problems through stochastic
  contraction, in: Proceedings of the IEEE/CVF Conference on Computer Vision
  and Pattern Recognition (CVPR), 2022, pp. 12413--12422.

\bibitem{xu2023restart}
Y.~Xu, M.~Deng, X.~Cheng, Y.~Tian, Z.~Liu, T.~Jaakkola, Restart sampling for
  improving generative processes, in: Advances in Neural Information Processing
  Systems, Vol.~36, 2023, pp. 76806--76838.

\bibitem{feng2019hypergraph}
Y.~Feng, H.~You, Z.~Zhang, R.~Ji, Y.~Gao, Hypergraph neural networks, in:
  Proceedings of the AAAI Conference on Artificial Intelligence, Vol.~33, 2019,
  pp. 3558--3565.

\bibitem{manduca2009projection}
A.~Manduca, L.~Yu, J.~D. Trzasko, N.~Khaylova, J.~M. Kofler, C.~M. McCollough,
  J.~G. Fletcher, Projection space denoising with bilateral filtering and {CT}
  noise modeling for dose reduction in {CT}, Medical Physics 36~(11) (2009)
  4911--4919.

\bibitem{xu2012low}
Q.~Xu, H.~Yu, X.~Mou, L.~Zhang, J.~Hsieh, G.~Wang, Low-dose {X}-ray {CT}
  reconstruction via dictionary learning, IEEE Transactions on Medical Imaging
  31~(9) (2012) 1682--1697.

\bibitem{zhang2025sparse}
C.~Zhang, K.~S. Yen, Sparse risk estimation and consistency-regulated
  unsupervised framework for efficient medical image denoising, Knowledge-Based
  Systems 330 (2025) 114646.

\bibitem{amir}
Z.~Yang, H.~Chen, Z.~Qian, Y.~Yi, H.~Zhang, D.~Zhao, B.~Wei, Y.~Xu, All-in-one
  medical image restoration via task-adaptive routing, in: International
  Conference on Medical Image Computing and Computer-Assisted Intervention
  (MICCAI), Springer, 2024, pp. 67--77.

\bibitem{lu2026allfreqnet}
Z.~Lu, X.~Hu, P.~Dong, Y.~Zha, J.~Yao, B.~Su, {AllFreqNet}: A unified
  frequency-enhanced all-in-one framework for medical image restoration,
  Knowledge-Based Systems 352 (2026) 116882.

\bibitem{xie2025reffusion}
D.~Xie, X.~Hu, H.~Xiao, Y.~Zhou, S.~Duan, Reffusion: Enhancement conditional
  diffusion framework with dual domain interaction transformer for image
  restoration, Knowledge-Based Systems 311 (2025) 112998.

\bibitem{meng2022sdedit}
C.~Meng, Y.~He, Y.~Song, J.~Song, J.~Wu, J.-Y. Zhu, S.~Ermon, {SDEdit}: Guided
  image synthesis and editing with stochastic differential equations, in:
  International Conference on Learning Representations (ICLR), 2022.

\bibitem{perez2018film}
E.~Perez, F.~Strub, H.~de~Vries, V.~Dumoulin, A.~Courville, {FiLM}: Visual
  reasoning with a general conditioning layer, in: Proceedings of the AAAI
  Conference on Artificial Intelligence, Vol.~32, 2018, pp. 3942--3951.

\bibitem{chen2025hypergraph}
C.~Chen, Y.~Wang, Y.~Zhang, N.~Zhang, H.~Feng, D.~Xu, Hypergraph neural network
  for remote sensing hyperspectral image super-resolution, Knowledge-Based
  Systems 321 (2025) 113755.

\bibitem{han2022vig}
K.~Han, Y.~Wang, J.~Guo, Y.~Tang, E.~Wu, Vision {GNN}: An image is worth graph
  of nodes, in: Advances in Neural Information Processing Systems, Vol.~35,
  2022, pp. 8291--8303.

\bibitem{han2023vihgnn}
Y.~Han, P.~Wang, S.~Kundu, Y.~Ding, Z.~Wang, Vision {HGNN}: An image is more
  than a graph of nodes, in: Proceedings of the IEEE/CVF International
  Conference on Computer Vision (ICCV), 2023, pp. 19878--19888.

\bibitem{chen2019glore}
Y.~Chen, M.~Rohrbach, Z.~Yan, S.~Yan, J.~Feng, Y.~Kalantidis, Graph-based
  global reasoning networks, in: Proceedings of the IEEE/CVF Conference on
  Computer Vision and Pattern Recognition (CVPR), 2019, pp. 433--442.

\bibitem{ronneberger2015unet}
O.~Ronneberger, P.~Fischer, T.~Brox, {U-Net}: Convolutional networks for
  biomedical image segmentation, in: International Conference on Medical Image
  Computing and Computer-Assisted Intervention (MICCAI), Springer, 2015, pp.
  234--241.

\bibitem{mayo2020}
T.~R. Moen, B.~Chen, D.~R. Holmes~III, X.~Duan, Z.~Yu, L.~Yu, S.~Leng, J.~G.
  Fletcher, C.~H. McCollough, Low-dose {CT} image and projection dataset,
  Medical Physics 48~(2) (2021) 902--911.

\bibitem{vanaarle2016astra}
W.~van Aarle, W.~J. Palenstijn, J.~Cant, E.~Janssens, F.~Bleichrodt,
  A.~Dabravolski, J.~De~Beenhouwer, K.~J. Batenburg, J.~Sijbers, Fast and
  flexible {X}-ray tomography using the {ASTRA} toolbox, Optics Express 24~(22)
  (2016) 25129--25147.

\bibitem{mccollough2017ldgc}
C.~H. McCollough, A.~C. Bartley, R.~E. Carter, B.~Chen, T.~A. Drees,
  P.~Edwards, D.~R. Holmes, A.~E. Huang, F.~Khan, S.~Leng, K.~L. McMillan,
  G.~J. Michalak, K.~M. Nunez, L.~Yu, J.~G. Fletcher, Low-dose {CT} for the
  detection and classification of metastatic liver lesions: Results of the 2016
  {Low Dose CT Grand Challenge}, Medical Physics 44~(10) (2017) e339--e352.

\bibitem{ji2022amos}
Y.~Ji, H.~Bai, C.~Ge, J.~Yang, Y.~Zhu, R.~Zhang, Z.~Li, L.~Zhang, W.~Ma,
  X.~Wan, P.~Luo, {AMOS}: A large-scale abdominal multi-organ benchmark for
  versatile medical image segmentation, in: Advances in Neural Information
  Processing Systems, Vol.~35, 2022, pp. 36722--36732.

\bibitem{biguri2016tigre}
A.~Biguri, M.~Dosanjh, S.~Hancock, M.~Soleimani, {TIGRE}: A {MATLAB-GPU}
  toolbox for {CBCT} image reconstruction, Biomedical Physics \& Engineering
  Express 2~(5) (2016) 055010.

\bibitem{wang2004ossart}
G.~Wang, M.~Jiang, Ordered-subset simultaneous algebraic reconstruction
  techniques ({OS-SART}), Journal of X-Ray Science and Technology 12~(3) (2004)
  169--177.

\bibitem{unad}
L.~Gu, W.~Deng, G.~Wang, {UNAD}: Universal anatomy-initialized noise
  distribution learning framework towards low-dose {CT} denoising, in: IEEE
  International Conference on Acoustics, Speech and Signal Processing (ICASSP),
  IEEE, 2024, pp. 1671--1675.

\bibitem{wganvgg}
Q.~Yang, P.~Yan, Y.~Zhang, H.~Yu, Y.~Shi, X.~Mou, M.~K. Kalra, Y.~Zhang,
  L.~Sun, G.~Wang, Low-dose {CT} image denoising using a generative adversarial
  network with {W}asserstein distance and perceptual loss, IEEE Transactions on
  Medical Imaging 37~(6) (2018) 1348--1357.

\bibitem{fsim}
L.~Zhang, L.~Zhang, X.~Mou, D.~Zhang, {FSIM}: A feature similarity index for
  image quality assessment, IEEE Transactions on Image Processing 20~(8) (2011)
  2378--2386.

\bibitem{nqm}
N.~Damera-Venkata, T.~D. Kite, W.~S. Geisler, B.~L. Evans, A.~C. Bovik, Image
  quality assessment based on a degradation model, IEEE Transactions on Image
  Processing 9~(4) (2000) 636--650.

\bibitem{vif}
H.~R. Sheikh, A.~C. Bovik, Image information and visual quality, IEEE
  Transactions on Image Processing 15~(2) (2006) 430--444.

\bibitem{isensee2021nnunet}
F.~Isensee, P.~F. Jaeger, S.~A.~A. Kohl, J.~Petersen, K.~H. Maier-Hein,
  nn{U}-{N}et: A self-configuring method for deep learning-based biomedical
  image segmentation, Nature Methods 18~(2) (2021) 203--211.

\end{thebibliography}

\end{document}